\documentclass[prd,twocolumn,preprintnumbers,showpacs,nofootinbib,superscriptaddress,notitlepage,floatfix]{revtex4-1}
\usepackage{hyperref}
\usepackage{graphicx}
\usepackage{amsmath}
\usepackage{amssymb}
\usepackage{slashed}
\usepackage{amsfonts}
\usepackage{xcolor}
\usepackage{multirow}
\usepackage[caption=false]{subfig}
\usepackage{array}
\usepackage{mwe}

\usepackage[utf8]{inputenc}
\usepackage[normalem]{ulem}

\DeclareMathOperator\re{Re}

\DeclareRobustCommand{\eq}[1]{Eq.~\eqref{eq:#1}}

\DeclareRobustCommand{\fig}[1]{Fig.~\ref{fig:#1}}

\DeclareRobustCommand{\app}[1]{App.~\ref{app:#1}}

\DeclareRobustCommand{\tbl}[1]{Table~\ref{tbl:#1}}

\DeclareRobustCommand{\eq}[1]{Eq.~(\ref{eq:#1})}

\newcommand{\MS}{{\overline{\mathrm{MS}}}}

\newcommand{\nn}{\nonumber}

\newcommand{\Tr}{\mathrm{Tr}}

\newcommand\bets{\begin{table*}}
\newcommand\eets[1]{\label{tb:#1}\end{table*}}

\begin{document}

\title{Parton Distributions from Boosted Fields in the Coulomb Gauge}


\author{Xiang Gao}
\affiliation{Physics Division, Argonne National Laboratory, Lemont, IL 60439, USA}

\author{Wei-Yang Liu}
\affiliation{Center for Nuclear Theory, Department of Physics and Astronomy, Stony Brook University, Stony Brook, New York 11794–3800, USA}

\author{Yong Zhao}
\affiliation{Physics Division, Argonne National Laboratory, Lemont, IL 60439, USA}

\begin{abstract}
We propose a new method to calculate parton distribution functions (PDFs) from lattice correlations of boosted quarks and gluons in the Coulomb gauge.
Compared to the widely used gauge-invariant Wilson-line operators, these correlations greatly simplify the renormalization thanks to the absence of linear power divergence. Besides, they enable access to larger off-axis momenta under preserved 3D rotational symmetry, as well as enhanced long-range precision that facilitates the Fourier transform.
We verify the factorization formula that relates this new observable to the quark PDF at one-loop order in perturbation theory.
Moreover, through a lattice calculation of the pion valence quark PDF, we demonstrate the aforementioned advantage and features of the Coulomb gauge correlation and show that it yields consistent result with the gauge-invariant method.
This opens the door to a more efficient way to calculate parton physics on the lattice.
\end{abstract}

\maketitle

One of the top quests in nuclear and particle physics nowadays is to understand the 3D internal structure of the proton.
Over the past five decades, high-energy scattering experiments at facilities including SLAC, COMPASS, HERA, Fermilab, LHC, Jefferson Lab and RHIC have provided state-of-the-art measurement of the proton parton distribution functions (PDFs)~\cite{ParticleDataGroup:2022pth}, which are 1D densities of quarks and gluons in their momentum fraction $x$, as well as the 3D and spin-dependent distributions. The future Electron-Ion Collider (EIC) will continue the endeavor with unprecedented precision~\cite{Accardi:2012qut,AbdulKhalek:2021gbh}.

The experimental pursuit of the proton 3D structure has also motivated its first-principles calculation from lattice quantum chromodynamics (QCD), a Euclidean formulation of quantum field theory. However, for a long time such efforts have been hindered by the real-time dependence of light-cone correlations that define the PDFs, which makes them not directly calculable on a Euclidean lattice with imaginary time.
About a decade ago, a breakthrough was made with the proposal of \textit{large-momentum effective theory} (LaMET)~\cite{Ji:2013dva,Ji:2014gla,Ji:2020ect}, which starts from the quasi-PDF (qPDF) defined as Fourier transform of an equal-time correlation at large proton momentum, and relates it to the PDF through power expansion and effective theory matching~\cite{Ji:2020byp}. Over the past years, among the other methods proposed~\cite{Liu:1993cv,Detmold:2005gg,Braun:2007wv,Chambers:2017dov,Radyushkin:2017cyf,Ma:2017pxb}, LaMET has made the most significant progress in the calculation of PDFs and pioneered the studies of 3D partonic structures~\cite{Ji:2020ect,Constantinou:2020hdm,Boussarie:2023izj}.

At the core of LaMET is the simulation of nonlocal bilinear operators which define the qPDFs~\cite{Ji:2013dva}. For example, the quark bilinear is $O_\Gamma(z)\equiv \bar{\psi}(z)\Gamma W(z,0) \psi(0)$, where $\Gamma$ is a Dirac matrix, and $W(z,0)$ is a spacelike Wilson line that connects $0$ to $z^\mu=(0,\vec{z})$ to make $O_\Gamma(z)$ gauge invariant. 
By construction $O_\Gamma(z)$ must approach the light-cone $t+|\vec{z}|=0$ under a Lorentz boost along the $\vec{z}$-direction, which can be achieved on the lattice by simulating a boosted hadron. One major challenge here is to reach large momentum which controls the power accuracy. 
So far the most widely used method to achieve high momenta is the momentum smearing technique~\cite{Bali:2016lva}.
However, to ensure a smooth Wilson line both $\vec{z}$ and the momentum $\vec{p}$ must be along one spatial axis, which leaves out all the off-axis directions that can be used to reach higher momenta~\footnote{A zigzag Wilson-line operator was studied in Ref.~\cite{Musch:2010ka}, where rotational symmetry was found to be weakly broken on smeared gauge configurations, but it is hard to quantify this error.}.
Another important issue is the renormalization of $O_\Gamma(z,a)$ under lattice regularization with spacing $a$, as it includes a linear power divergence $\propto \exp(-\delta m(a)|\vec{z}|)$ with $\delta m(a) \!\sim \!1/a$, which originates from the Wilson-line self-energy~\cite{Dorn:1986dt,Maiani:1991az,Ji:2017oey,Ishikawa:2017faj,Green:2017xeu}. In order to calculate the $x$-dependence of PDFs, such a divergence must be subtracted at all $\vec{z}$~\cite{Ji:2020brr}, and a nontrivial matching onto the $\MS$ scheme~\cite{Holligan:2023rex,Zhang:2023bxs} is required to cancel the associated renormalon $\exp(-m_0|\vec{z}|)$ with $m_0\!\sim\! \Lambda_{\rm QCD}$~\cite{Bali:2013pla}, thus eliminating the ${\cal O}(\Lambda_{\rm QCD}/|\vec{p}|)$ power correction~\cite{Zhang:2023bxs}. 

In this work we propose to calculate the PDFs from pure quark and gluon correlations in the Coulomb gauge (CG), within the framework of LaMET. The qPDF defined from such a correlation falls into the same universality class~\cite{Hatta:2013gta,Ji:2020ect} as the gauge-invariant (GI) qPDF, since they both approach the PDF under an infinite Lorentz boost. 
Without the Wilson line, the CG correlation is free from the linear divergence and renormalon, which greatly simplifies the renormalization. 
Besides, the computation and storage cost for Wilson lines can be reduced in lattice simulations, and one can reach larger off-axis momenta by taking advantage of the 3D rotational symmetry of CG. Moreover, since the renormalization factor is independent of $z$, the exponential decaying bare correlation at large $|\vec{z}|$ is unaffected, which will enhance the precision and facilitate the Fourier transform.
At last, one can do the momentum smearing~\cite{Izubuchi:2019lyk,Gao:2020ito} in the CG and compute both GI and CG qPDFs simultaneously, as they share the same quark propagators.

In the following, we first introduce the CG quark qPDF and verify its LaMET matching to the PDF at next-to-leading order (NLO). Next, we calculate the pion valence quark PDF using both methods with the same lattice setup, with the inclusion of a large off-axis momentum for the CG qPDF. 
We demonstrate the aforementioned efficiencies of the CG qPDF and confirm the consistency of both methods. 
Finally, we discuss the broader application of CG correlations in calculating parton physics.

\paragraph*{Definition.} The idea of using CG is not new, as it was first proposed within the LaMET framework for calculating the gluon helicity contribution to the proton spin~\cite{Ji:2013fga,Hatta:2013gta,Ji:2014lra,Yang:2016plb}.
The CG quark qPDF is defined as~\footnote{While the CG correlation was examined for the pion wave function in Ref.~\cite{Gupta:1993vp}, it was not adopted as a formal approach there.}
\begin{align}
    \tilde f(x,P^z,\mu) &= P^z\int_{-\infty}^\infty \frac{dz}{2\pi} e^{i x P^z z} \tilde h(z,P^z,\mu)\,,\\
    \tilde h(z,P^z,\mu) &= {1\over 2P^t}\langle P| \bar{\psi}(z)\gamma^t \psi(0)\Big|_{\vec{\nabla}\cdot \vec{A}=0}|P\rangle\,,
\end{align}
where $z^\mu=(0,0,0,z)$, $|P\rangle$ is a hadron state with $P^\mu=(P^t,0,0,P^z)$ normalized to $\langle P|P\rangle = 2P^t\delta^{(3)}(0)$, and $\mu$ is the $\MS$ scale. The GI qPDF follows a similar definition except that the quark correlator is replaced with $O_{\gamma^t}(z)$. The CG condition $\vec{\nabla}\cdot \vec{A}=0$ is fixed so that the quark correlation can have a nonvanishing matrix element. Due to 3D rotational invariance, the correlator and hadron momentum can be oriented to any spatial direction.

Meanwhile, the quark PDF $f(x,\mu)$ is defined as
\begin{align}
    f(x,\mu) &= \int_{-\infty}^\infty \frac{d\lambda}{2\pi} e^{-i\lambda x} h(\lambda,\mu)\,,\\
    h(\lambda,\mu) &= \frac{1}{2P^+} \langle P| \bar{\psi}(\xi^-)W(\xi^-,0) \gamma^+ \psi(0)|P\rangle\,,
\end{align}
where $\lambda=P^+\xi^-$ and $\xi^- = (t- z)/\sqrt{2}$.  Under an infinite Lorentz boost, the CG reduces to the light-cone gauge $A^+=(A^t+A^z)/\sqrt{2}=0$ with a proper boundary condition, where $W(\xi^-,0)=P\exp\big[-ig\int_0^{\xi^-}d\eta^-\ A^+(\eta^-)\big]$ vanishes, so the qPDF becomes equivalent to the PDF.

\paragraph*{Factorization.} According to LaMET~\cite{Ji:2020ect}, when $P^z\gg \Lambda_{\rm QCD}$ the CG qPDF can be perturbatively matched onto the PDF through a factorization formula~\cite{Izubuchi:2018srq},
\begin{align}\label{eq:fact}
	\tilde f(x,P^z,\mu) &= \int {dy\over |y|}\ C\Big({x\over y}, {\mu\over |y|P^z}\Big) f(y,\mu) \nn\\
	&\qquad\qquad + {\cal O}\Big({\Lambda_{\rm QCD}^2\over x^2P_z^2},{\Lambda_{\rm QCD}^2\over(1-x)^2P_z^2}\Big)\,,
\end{align}
where $C$ is the matching coefficient, and the power corrections are suppressed by the active and spectator parton momenta.
The factorization is based on effective field theory principles and can be proved using the Feynman diagram analysis for the GI qPDF~\cite{Ma:2014jla,Ji:2020ect}, which will be an object for future study.

By calculating the NLO corrections to the quark CG qPDF and PDF in a free quark state, we find out that their collinear divergences are identical~\cite{Xiong:2013bka}, which confirms \eq{fact} at the same order. The $\MS$ matching coefficient is obtained as a series in the strong coupling $\alpha_s$,
\begin{align}\label{eq:qpdf_match}
	C\big(\xi, {\mu\over p^z}\big) &= \delta(\xi\!-\!1) + {\alpha_sC_F\over 2\pi}C^{(1)}\big(\xi,{\mu\over p^z}\big) + {\cal O}(\alpha_s^2)\,,
\end{align}
where $C_F=4/3$.
At NLO,
\begin{align}
	&C^{(1)}\big(\xi,{\mu\over p^z}\big) = C^{(1)}_{\rm ratio}\big(\xi,{\mu\over p^z}\big)\\
	& + {1\over 2|1-\xi|} +\delta(1-\xi) \left[-{1\over 2}\ln {\mu^2\over 4p_z^2} +{1\over2} - \int_0^1 d\xi' {1\over 1-\xi'}\right]\,,\nn
\end{align}
where
\begin{align}
    &C^{(1)}_{\rm ratio}\big(\xi,{\mu\over p^z}\big) = \left[P_{qq}(\xi)\ln{4p_z^2\over\mu^2} + \xi - 1\right]_{+(1)}^{[0,1]} \\
    & \quad + \Bigg\{P_{qq}(\xi) \Big[\text{sgn}(\xi)\ln|\xi| + \text{sgn}(1\!-\!\xi)\ln|1\!-\!\xi|\Big] \!+\! \text{sgn}(\xi)\nn\\
    &\quad + {3 \xi-1\over \xi-1} \frac{\tan^{-1}\left(\sqrt{1-2 \xi}/|\xi|\right)}{\sqrt{1-2 \xi}} - {3\over 2|1-\xi|} \Bigg\}_{+(1)}^{(-\infty, \infty)}\nn
\end{align}
corresponds to the ratio scheme~\cite{Orginos:2017kos} that satisfies particle number conservation. Here $P_{qq}(\xi)=(1+\xi^2)/(1-\xi)$, and the plus functions are defined on a domain $D$ as
\begin{align}
	\left[g(x)\right]^D_{+(x_0)} &= g(x) - \delta(x-x_0)\int_D dx'\ g(x')\,.
\end{align}
Note that $C^{(1)}_{\rm ratio}$ is analytical at $\xi=1/2$ despite its form.

With a double Fourier transform of \eq{fact}~\cite{Izubuchi:2018srq}, we also derive a short-distance factorization (SDF):
\begin{align}\label{eq:sdf}
	\tilde h(z,\!P^z,\!\mu) &\!=\! \int\! du\ {\cal C}(u,\!z^2\mu^2) h(u \tilde\lambda,\!\mu) \!+\! {\cal O}(z^2\Lambda_{\rm QCD}^2)\,,
\end{align}
where $\tilde\lambda = zP^z$. 
Like \eq{qpdf_match}, the NLO coefficient is
\begin{align}
	{\cal C}^{(1)}(u,z^2\mu^2) &= {\cal C}^{(1)}_{\rm ratio}(u,z^2\mu^2)  \!+\! \delta(1\!-\!u)\big({1\over2}-{\mathbf{L}_z\over2}\big)\,,
\end{align}
where $\mathbf{L}_z=\ln\big(z^2\mu^2e^{2\gamma_E}/ 4\big)$, and
\begin{align}
    &{\cal C}^{(1)}_{\rm ratio}(u,z^2\mu^2)    = \left[-P_{qq}(u)\mathbf{L}_z - {4\ln(1\!-\!u)\over 1\!-\!u} + 1-u\right]_{+(1)}^{[0,1]} \nn\\
    & + \Bigg[ {3 u-1\over u-1} \frac{\tan^{-1}\left(\sqrt{1\!-\!2 u}/|u|\right)}{\sqrt{1-2 u}} \!-\! {3\over |1-u|} \Bigg]_{+(1)}^{(-\infty, \infty)}\,.
\end{align}

In contrast to the matching for GI correlations, where $u$ is limited to $[-1,1]$~\cite{Radyushkin:2017cyf,Ji:2017rah}, the ${\cal C}^{(1)}(u,z^2\mu^2)$ here is nonzero for $u<0$ and $u>1$. As a result, the Mellin moments of ${\cal C}^{(1)}(u,z^2\mu^2)$ are divergent except for the lowest one, indicating that the Wilson coefficients in the operator product expansion of $\tilde h(z,P^z,\mu)$ are functions not only of $z^2$ and $\mu^2$ but also of $P_z^2$. This feature is distinct from the GI case~\cite{Izubuchi:2018srq} and will be further studied.

\paragraph*{Numerical implementation.}
To test the CG method, we calculate the pion valence quark PDF on a gauge ensemble in 2+1 flavor QCD 
generated by the HotQCD collaboration~\cite{HotQCD:2014kol} with
Highly Improved Staggered Quarks~\cite{Follana:2006rc}, where the lattice spacing $a=0.06$ fm and volume $L_s^3\times L_t=48^3 \times 64$. We use tadpole-improved clover Wilson valence fermions 
on the hypercubic (HYP) smeared~\cite{Hasenfratz:2001hp} gauge background, with a valence pion mass $m_\pi=300$ MeV. To improve the signal of boosted pions at $\vec{p}=(2\pi)/(L_sa)\vec{n}$, we utilize the momentum-smeared~\cite{Bali:2016lva} pion source with optimized quark boost $\vec{k}$~\cite{Izubuchi:2019lyk,Gao:2020ito}.
We employ 109 gauge configurations and perform multiple exact and sloppy Dirac operator inversions on each of them using All Mode Averaging~\cite{Shintani:2014vja}. We use $\vec{n}=(0,0,0)$, two on-axis $\vec{n}=(0,0,n_z)$ with $n_z=4,5$ corresponding to $|\vec{p}|=1.72, 2.15$ GeV, and one off-axis $\vec{n}=(3,3,3)$ which corresponds to $|\vec{p}|=2.24$ GeV. Three time separations $t_s/a=8,10,12$ are computed to eliminate the excited-state contamination. Since the quark propagators are the same, we calculate the GI qPDF with 1-step HYP-smeared Wilson lines and the CG qPDF during contraction at no additional cost. More details of the statistics are shown in \tbl{setup}.

\begin{table}
\centering
\begin{tabular}{c|c|c|c|c}
\hline
\hline
$|\vec{p}|$ (GeV)&$\vec{n}$ & $\vec{k}$ & $t_s/a$ & (\#ex,\#sl) \\
\hline
 0   &(0,0,0) & (0,0,0) & 8,10,12 & (1, 16) \\
\hline
     && & 8       & (1, 32) \\
 1.72&(0,0,4) & (0,0,3) & 10      & (3, 96) \\
     && & 12      & (8, 256) \\
\hline
     && & 8       & (2, 64) \\
 2.15&(0,0,5) & (0,0,3) & 10      & (8, 256) \\
     && & 12      & (8, 256) \\
 \hline
     & &  & 8       & (2, 64) \\
 2.24&(3,3,3) & (2,2,2) & 10      & (8, 256) \\
     && & 12      & (8, 256) \\
\hline
\hline
\end{tabular}
\caption{Details of lattice setup, where $\vec{p}=(2\pi)/(L_sa)\vec{n}$, $\vec{k}$ is the momentum-smearing parameter~\cite{Bali:2016lva}, $t_s$ is the source-sink separation, and (\#ex,\#sl) are the numbers of exact and sloppy inversions per configuration.}
\label{tbl:setup}
\end{table}

For a 4D lattice of spatial volume $V$, we fix QCD in the CG by finding the gauge transformation $\Omega$ of link variables $U_i(t,\vec{x})$ that minimizes the criterion~\cite{Davies:1987vs,Hudspith:2014oja}
\begin{align}
\label{eq:gribov}
  F[U^\Omega]
 & = \frac{1}{9V} \sum_{\vec{x}} \sum_{i=1,2,3} \big[-\re\Tr\ U^\Omega_i(t,\vec{x})\big]
\end{align}
per time slice. Each gauge fixing takes 600 sweeps and reaches a precision of $<\!10^{-7}$. Though imprecise fixing and the presence of Gribov copies~\cite{Gribov:1977wm,Singer:1978dk} can affect gauge-variant correlations, they most likely contribute to the statistical noise with our algorithm~\cite{Giusti:2001xf}. In our simulation, increasing the statistics reduces the overall statistical error (see \app{Gribov}), suggesting that the Gribov noise is not important~\cite{Burgio:2012ph}. Besides, lattice studies of the SU(2) gluon propagator in the Landau gauge~\cite{Maas:2017csm} and CG~\cite{Burgio:2016nad} show that Gribov copies only affect the far infrared region $\lesssim 0.2$ GeV, which implies that they might have negligible impact on the QCD PDF at $2x|\vec{p}|\gg 0.2$ GeV where LaMET is reliable~\cite{Zhang:2023bxs}.
A further study of the effect of Gribov copies on the CG qPDF will be carried out in the future.

Using an off-axis momentum $\vec{n}=(n_x,n_y,n_z)$, one can achieve the same $|\vec{n}|$ with less oscillatory modes $n_{x,y,z}$. 
Compared to $\vec{n}=(0,0,5)$, we observe in $\vec{n}=(3,3,3)$ about 20\% increase in the signal-to-noise ratios of both two-point and three-point correlations at $t_s/a \le 10$ (see \app{bare}). Besides, we also find that 3D rotational symmetry is precisely maintained in the case of CG correlations, whereas it is broken to some extent in the GI case (see \app{bare2}).

\begin{figure}
    \centering
    \includegraphics[width=0.4\textwidth]{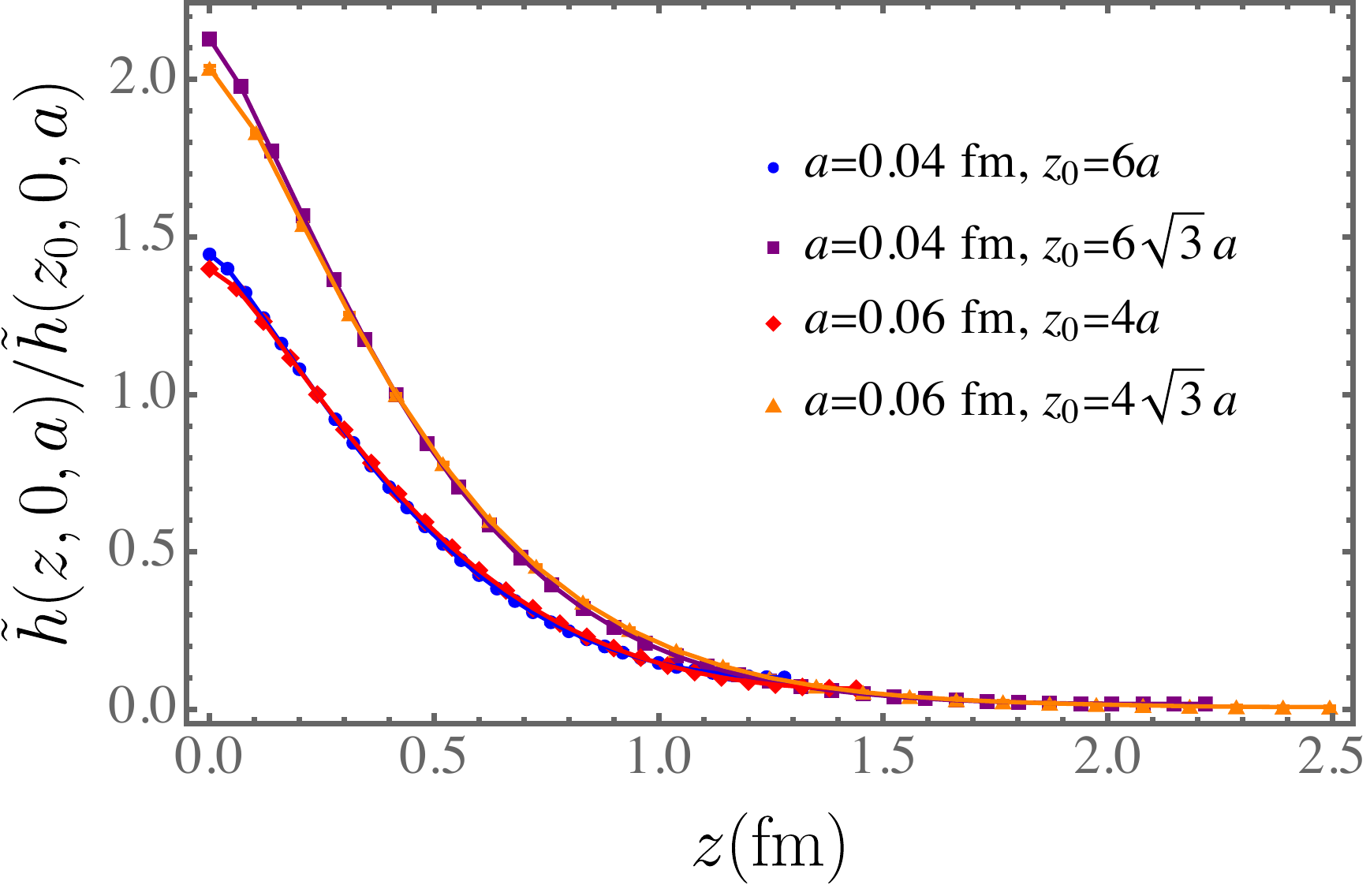}
    \caption{Comparison of CG ratios $\tilde{h}(z,0,a)/\tilde{h}(z_0,0,a)$ at $a=0.04$ and $0.06$ fm, with $z_0$ fixed in the physical unit. For more examples, we choose $\vec{z}$ to be along $\vec{n}=(0,0,1)$ for blue and red points and $\vec{n}=(1,1,1)$ for purple and orange points.}
    \label{fig:CGRnm}
\end{figure}

With the bare matrix elements, our next step is renormalization. Since QCD has been proven renormalizable in CG~\cite{Zwanziger:1998ez,Baulieu:1998kx,Niegawa:2006ey} without linear divergence~\cite{Niegawa:2006hg}, renormalization of the quark correlator is simply multiplicative,
\begin{align}\label{eq:multp}
	\bar{\psi}_B(z) \Gamma \psi_B(0) &= Z_\psi(a) \left[\bar{\psi}(z) \Gamma \psi(0)\right]_R\,,
\end{align}
where $B$ and $R$ stand for bare and renormalized quantities, respectively, and $Z_\psi$ is the quark wave function renormalization factor. This has been verified for the quark propagator on the lattice~\cite{Burgio:2012ph}. For hadronic matrix elements, the ratio $\tilde{h}(z,0,a)/\tilde{h}(z_0,0,a)$ should have a continuum limit according to \eq{multp}. On a finer HotQCD ensemble~\cite{HotQCD:2014kol} with $a=0.04$ fm and $L_s^3\times L_t=64^3 \times 64$, which was also used in Refs.~\cite{Gao:2020ito,Gao:2021dbh}, we compute the $\vec{p}=0$ matrix elements using 40 configurations with 1 exact and 16 sloppy inversions on each of them. The ratios at two lattice spacings are shown in \fig{CGRnm}.
As one can see, the results agree extremely well (except for those at $|\vec{z}|\sim a$ where discretization effects become important), thus proving the absence of linear divergence.

\begin{figure}
    \centering
    \includegraphics[width=0.9\columnwidth]{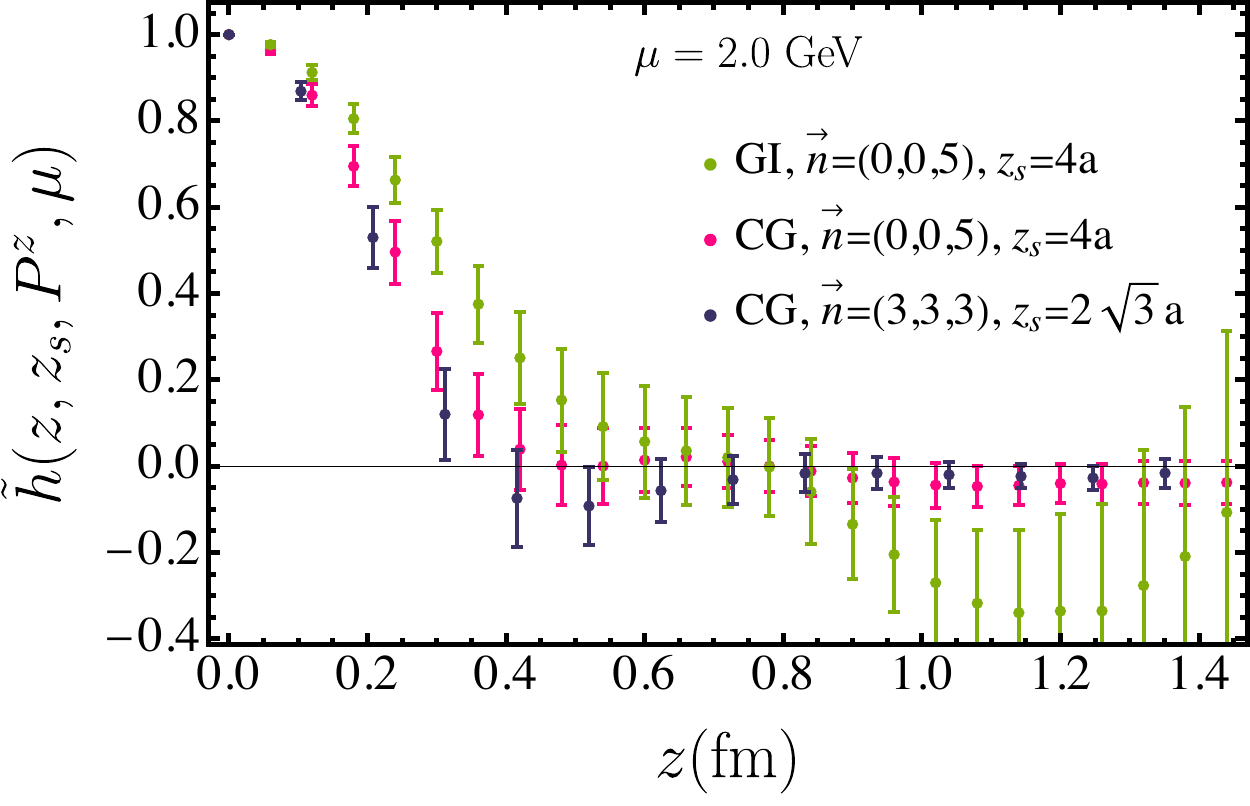}
    \caption{CG and GI correlations in the hybrid scheme at on-axis momentum $2.15$ GeV with $\vec{n}=(0,0,5)$ and off-axis momentum $2.24$ GeV with $\vec{n}=(3,3,3)$.}
    \label{fig:hbd}
\end{figure}

Then we do renormalization in the hybrid scheme~\cite{Ji:2020brr}, 
\begin{align}\label{eq:hbd}
	&\tilde h(z,z_s,P^z,\mu) \!=\! N\big[\tilde h(z,P^z,a) / \tilde h(z,0,a)\big] \theta(z_s\!-\!|z|) \\
	&~+ N e^{(\delta m+ m_0)(|z|-z_s)} \big[\tilde h(z,P^z,a) / \tilde h(z_s,0,a)\big] \theta(|z|\!-\!z_s)\,,\nn
\end{align}
where $N=\tilde{h}(0,0,a)/\tilde{h}(0,P^z,a)$, and $z_s=4a$ and $2\sqrt{3}a$ for on- and off-axis momenta, respectively. For the GI correlation, $\delta m$ is the same as that in Ref.~\cite{Gao:2021dbh}, and $m_0(\mu)$ is fitted with the leading-renormalon resummation (LRR) approach under large-$\beta_0$ approximation~\cite{Holligan:2023rex,Zhang:2023bxs}.
A precise determination of $\delta m$ and $m_0$ typically requires multiple fine lattice spacings~\cite{Gao:2021dbh,LatticePartonCollaborationLPC:2021xdx,Zhang:2023bxs}.
In contrast, for the CG correlation $\delta m = m_0=0$, which does not need extra calculation, thus greatly simplifying the renormalization and eliminating related systematics.
Moreover, thanks to the absence of linear renormalon, the power corrections in the factorization formula \eq{fact} starts at the quadratic order, which is good for the convergence in $P^z$ after matching.
\fig{hbd} compares the hybrid-scheme CG and GI correlations. Both fall close to zero at large $z$, but the errors in the GI case are significantly larger due to the exponential enhancement by the subtraction of $\delta m$.
Next, we Fourier transform the correlations to obtain the qPDFs. The discrete data are interpolated with a cubic polynomial, whose uncertainty is small compared to the other systematics. For the GI correlation, we extrapolate to $z=\infty$ with a physically motivated model $e^{-m|z|}/\tilde \lambda^d$~\cite{Gao:2021dbh}, which mainly affects the small-$x$ region. Meanwhile, thanks to the simple renormalization, the extrapolation has much less impact on the CG qPDF as both the central value and error of the correlation remain small at large $z$.

\begin{figure}
    \centering
    \includegraphics[width=0.9\columnwidth]{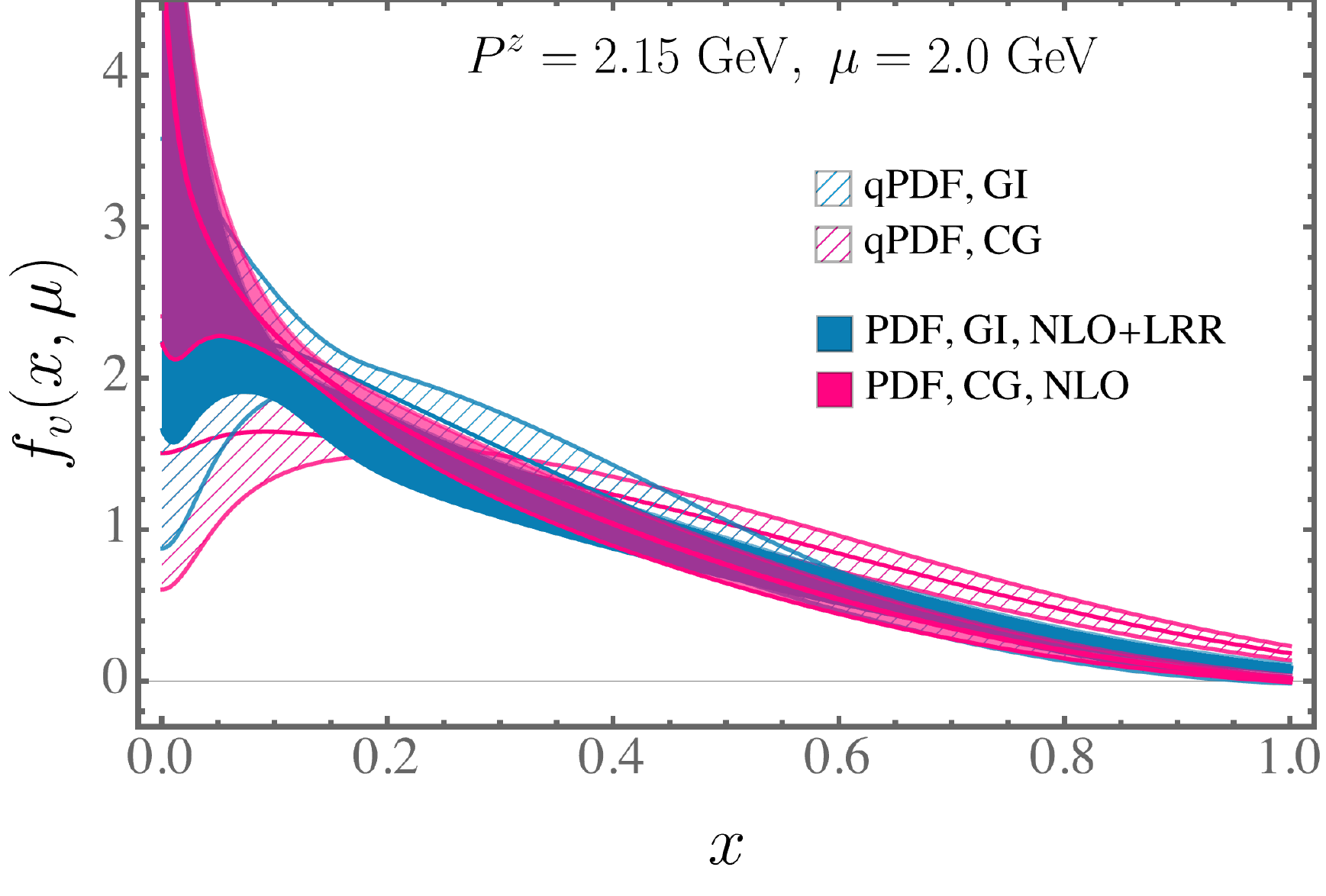}
    \caption{Comparison of the CG and GI qPDFs before and after matching at $P^z=2.15$ GeV and $\mu=2.0$ GeV. The error band only includes the statistical uncertainty.}
    \label{fig:match}
\end{figure}

Now we match the qPDFs to the PDF with NLO coefficient for the GI qPDF~\cite{Ji:2020brr}, which in the CG case is
\begin{align}
    C^{(1)}(\xi,z_s,p^z,\mu) &= C^{(1)}_{\rm ratio}\big(\xi,{\mu\over p^z}\big)\\
    &  - \left[ {{\rm Si}[(1-\xi)z_sp^z]\over \pi(1-\xi)}-{1\over2 |1-\xi|}\right]_{+(1)}^{(-\infty, \infty)}\,,\nn
\end{align}
where ${\rm Si}(\lambda)=\int_0^\lambda dt\ \sin t/t$. \fig{match} compares the CG and GI qPDFs before and after matching. 
Despite noticeable difference between the qPDFs, the matched results agree within errors at $x\gtrsim 0.2$. Such convergence is also observed on each of the 200 qPDF bootstrap samples (see \app{conv}), which clearly demonstrates the consistency between the CG and GI methods. In addition, we also find agreement between them using the ratio scheme and SDF in coordinate space~\cite{Orginos:2017kos}, as detailed in \app{sdf}.

\begin{figure}
    \centering
    \includegraphics[width=0.9\columnwidth]{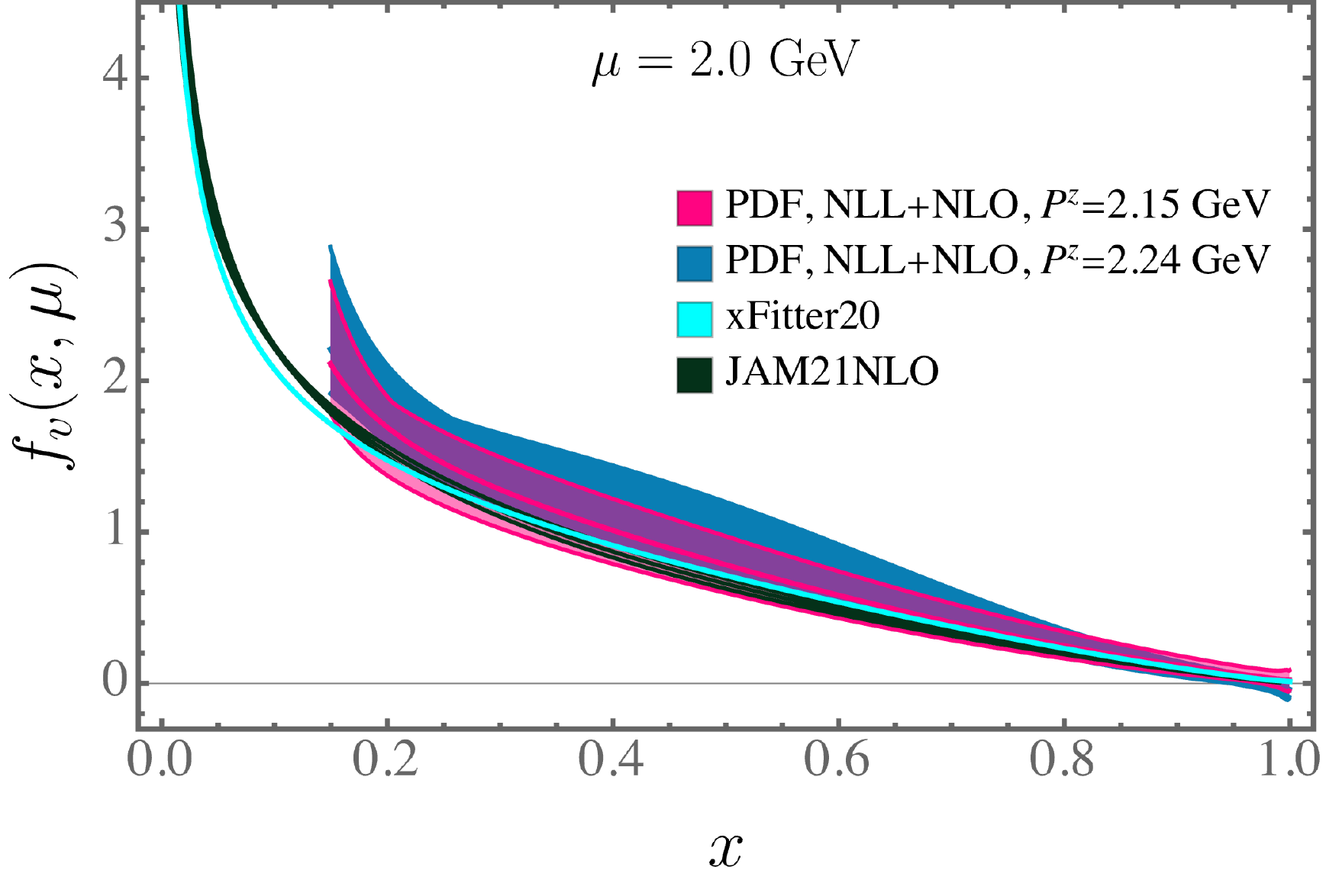}
    \caption{PDFs from the CG method at $|\vec{p}|=2.15$ and 2.24 GeV with NLO matching and NLL evolution, compared to \texttt{xFitter20}~\cite{Novikov:2020snp} and JAM21NLO~\cite{Barry:2021osv} fits. The lattice error bands include statistical and scale variation uncertainties.}
    \label{fig:evo}
\end{figure}

Finally, we conclude the analysis of CG qPDFs by resumming the small-$x$ logarithms through PDF evolution~\cite{Gao:2021hxl,Su:2022fiu}, while the resummation of large-$x$ logarithms~\cite{Gao:2021hxl,Ji:2023pba} is postponed. \fig{evo} shows the results at on-axis and off-axis momenta $|\vec{p}|=2.15$ and $2.24$ GeV, respectively, which are compared to the recent global fits by \texttt{xFitter20}~\cite{Novikov:2020snp} and JAM21NLO~\cite{Barry:2021osv}. The error has included scale variation, which is estimated by setting $\mu=2\kappa x|\vec{p}|$ with $\kappa=\sqrt{2},1,1/\sqrt{2}$ in the matching and evolving the results to $\mu=2.0$ GeV at next-to-leading-logarithmic (NLL) order. The resummation has a huge impact at $x\lesssim 0.2$ where the parton momentum $x|\vec{p}|$ approaches the Landau pole. For $x>0.2$, the lattice results agree with the global fits, though with larger errors.
More analysis details are provided in the Appendices.

In summary, we have proposed a new method to calculate the PDF from CG correlations within the LaMET framework. The factorization relation between the CG qPDF and PDF has been verified at NLO. 
With an exploratory lattice calculation, we show that the CG correlation is free from linear divergence and renormalon, which greatly simplifies the renormalization, and that it yields consistent results with the GI method. It also enables access to larger off-axis momenta under 3D rotational symmetry and enhances long-range precision, both contributing to more efficient lattice analysis.
There is still ample room for improvement, as we can increase the statistics and use multiple lattice ensembles for physical extrapolations.
Besides, the large-$x$ resummation is similar to the GI case and will be developed in the future.

Last but not least, the CG method can be applied to broader parton physics like generalized parton distributions and transverse-momentum distributions (TMDs), which are more computationally demanding than the PDFs. Especially, the TMD calculation will benefit from the absence of staple-shaped Wilson lines with significant noise reduction and greatly simplified renormalization~\cite{Constantinou:2019vyb,Ebert:2019tvc,Shanahan:2019zcq,Green:2020xco,Ji:2021uvr,Zhang:2022xuw,Alexandrou:2023ucc}. Since the boosted quarks in a physical gauge capture the correct collinear partonic degrees of freedom, their 3D correlation should be factorizable to the TMD~\cite{Ebert:2018gzl,Ebert:2019okf,Ji:2019sxk,Ji:2019ewn,Ebert:2022fmh}, which will be studied in a future work.

\begin{acknowledgments}

We thank Jack Holligan, Xiangdong Ji, Swagato Mukherjee, Peter Petreczky and Rui Zhang for valuable communications.
This material is based upon work supported by the U.S. Department of Energy, Office of Science, Office of Nuclear Physics through Contract No.~DE-AC02-06CH11357 and No.~DE-FG-88ER40388, and within the frameworks of Scientific Discovery through Advanced Computing (SciDAC) award \textit{Fundamental Nuclear Physics at the Exascale and Beyond} and the Quark-Gluon Tomography (QGT) Topical Collaboration, under contract no.~DE-SC0023646. 
YZ is also partially supported by the 2023 Physical Sciences and Engineering (PSE) Early Investigator Named Award program at Argonne National Laboratory.
We gratefully acknowledge the computing resources provided on \textit{Swing}, a high-performance computing cluster operated by the Laboratory Computing Resource Center at Argonne National Laboratory.
This research also used awards of computer time provided by the INCITE program at Argonne Leadership Computing Facility, a DOE Office of Science User Facility operated under Contract No.~DE-AC02-06CH11357. 
The computation of the correlators was carried out with the \texttt{Qlua} software suite~\cite{qlua}, which utilized the multigrid solver in \texttt{QUDA}~\cite{Clark:2009wm,Babich:2011np}.

\end{acknowledgments}
 
 \appendix

\section{Simulation of bare matrix elements}
\label{app:sim}

\subsection{Two-point and three-point functions}\label{app:bare}

\begin{figure}
    \centering
    \includegraphics[width=0.4\textwidth]{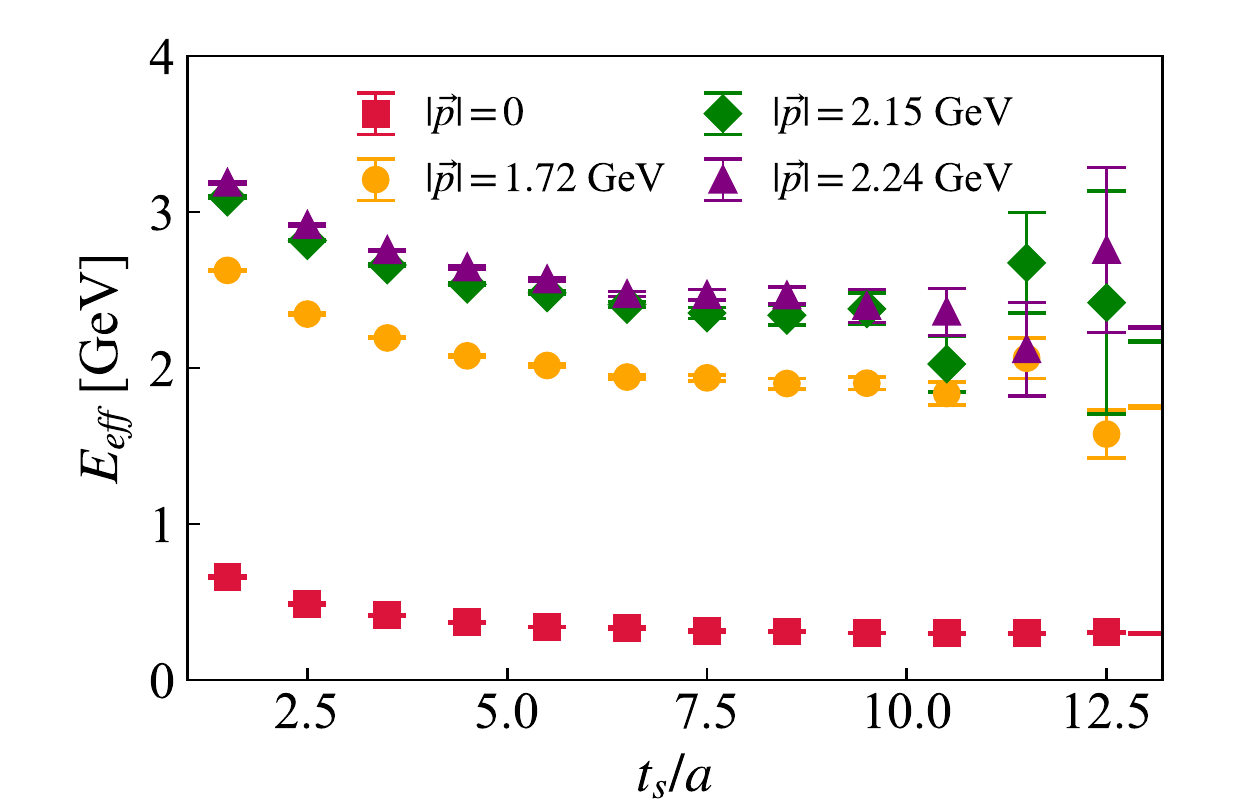}
    \caption{The effective mass evaluated from two-point functions as a function of $t_s$ are shown. The short colored lines on the right side are estimated from the disperion relation $E=\sqrt{\vec{p}^2+m_\pi^2}$ with $m_\pi$ = 300 MeV.}
    \label{fig:effmass}
\end{figure}

To determine the bare matrix elements of pion ground state, we first need the two-point functions $C_{2pt}(t_s;\vec{p})$ which will provide energy spectrum created by the pion source and corresponding overlap amplitudes~\cite{Izubuchi:2019lyk}. We utilize the Guassian momentum-smeared sources to improve the signal of boosted pion at momentum $\vec{p}=(2\pi)/(L_sa)\vec{n}$~\cite{Bali:2016lva}. We use $\vec{n}=(0,0,0)$, two on-axis $\vec{n}=(0,0,n_z)$ with $n_z=4,5$ which correspond to $|\vec{p}|=1.72$ and $2.15$ GeV, and one off-axis $\vec{n}=(3,3,3)$ which corresponds to $|\vec{p}|=2.24$ GeV. The optimized quark boost parameters and statistics are shown in \tbl{setup}. In \fig{effmass}, we show the effective mass evaluated from two-point functions as a function of time separation $t_s$. At $t_s\gtrsim 10a$ the effective mass, dominated by the pion ground state, agree with the short colored lines on the right side estimated from the dispersion relation $E=\sqrt{\vec{p}^2+m_\pi^2}$ with $m_\pi$ = 300 MeV. In addition, the results of $|\vec{p}|=2.24$ GeV case appear to be more stable than the $2.15$ GeV case.

\begin{figure}
    \centering
    \includegraphics[width=0.4\textwidth]{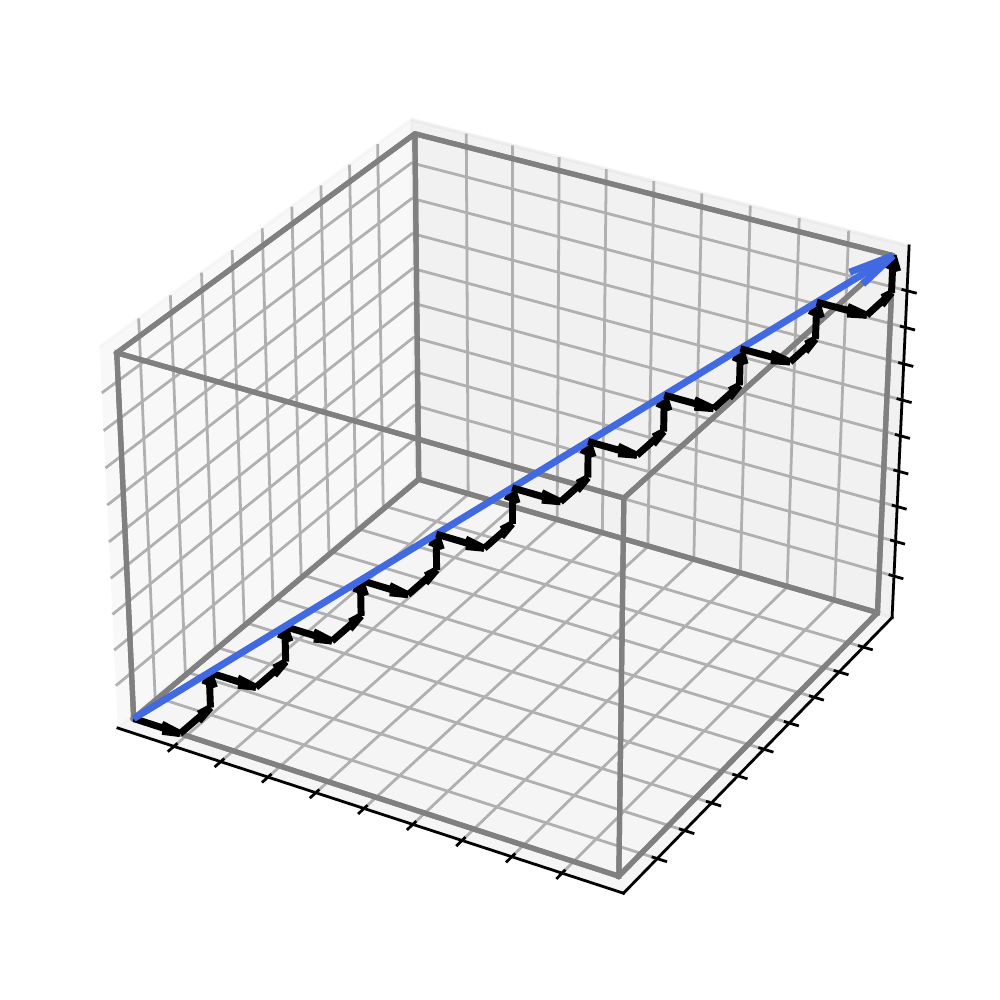}
    \caption{The black arrows are the zigzag Wilson lines for GI matrix elements with off-axis momentum.}
    \label{fig:zigzag}
\end{figure}

\begin{figure}
    \centering
    \includegraphics[width=0.4\textwidth]{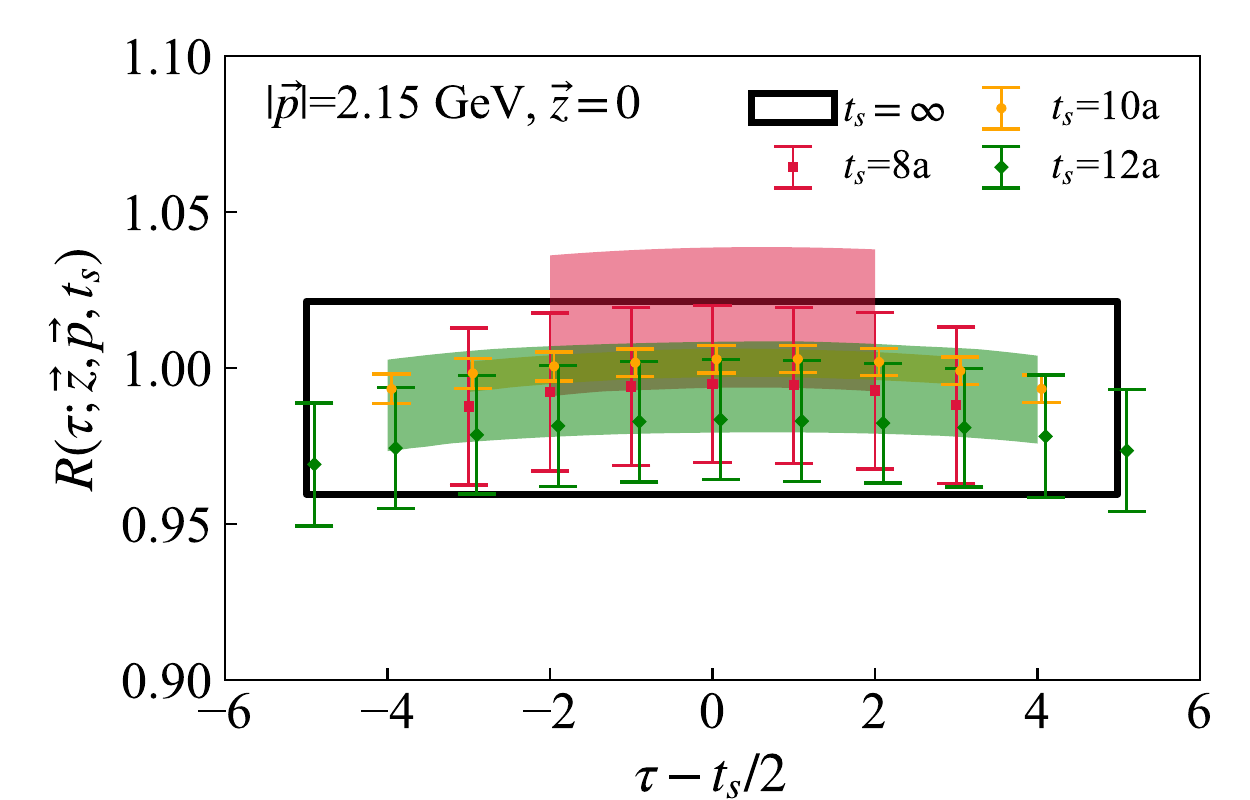}
    \includegraphics[width=0.4\textwidth]{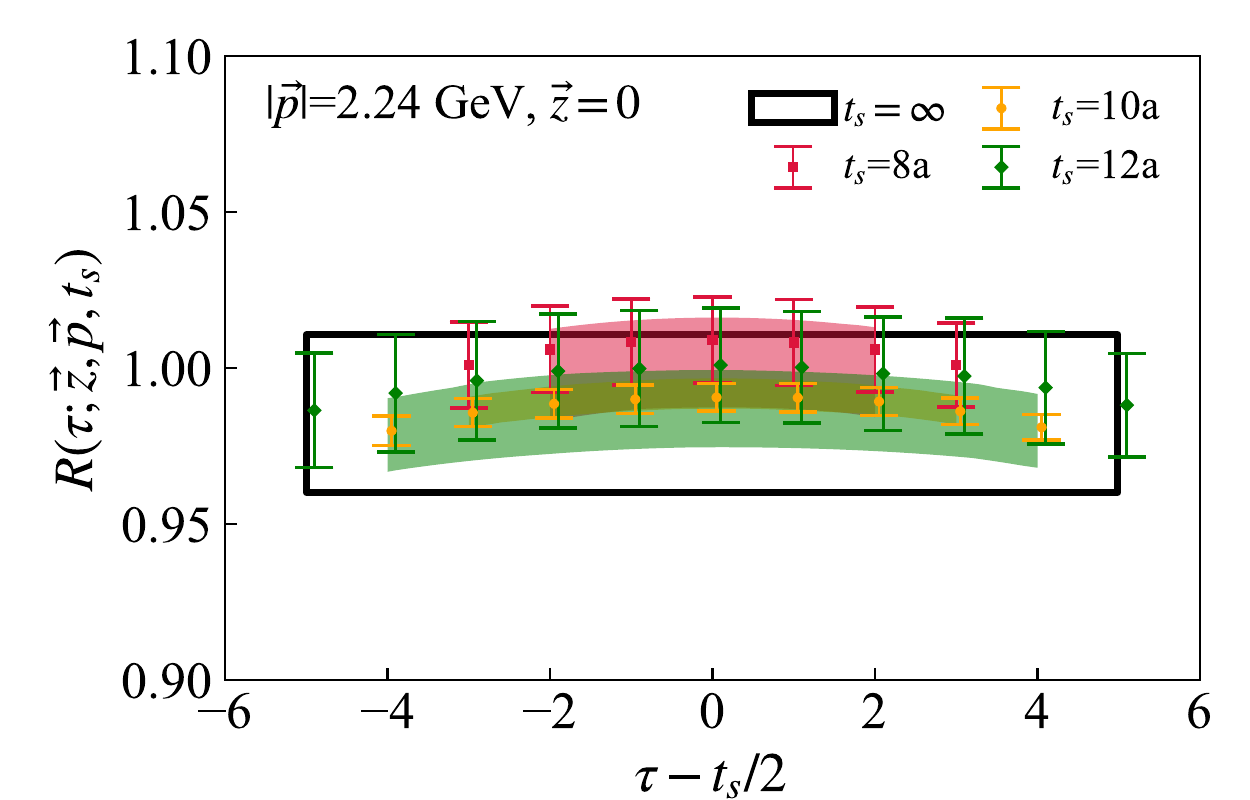}
    \caption{The $C_{3pt}/C_{2pt}$ ratios $R(\tau;\vec{z},\vec{p},t_s)$ at $\vec{z}=\vec{0}$ of $|\vec{p}|$ = 2.15 (upper panel) and 2.24 GeV (low panel) cases are shown.}
    \label{fig:ratiofit}
\end{figure}

To extract the the qPDF matrix elements, we need to compute the three-point functions $C_{3pt}(\tau,t_s;\vec{p})$. For the case of CG qPDF, we directly do the contraction of the quark propagators without Wilson line, using spatial separation $\vec{z}$ along the direction $\vec{n}$. As for the case of GI qPDF, we use straight Wilson lines $\vec{z}=(0,0,z_3)$ for on-axis momentum and zigzag Wilson lines for the off-axis momentum~\cite{Musch:2010ka}, as shown in \fig{zigzag}. As a result, the distance of a off-axis separation $\vec{z}=\{b,b,b\}$ is $|\vec{z}|=\sqrt{3}b$, while the total length of the Wilson line is $l=3b$. We construct the ratios $R(\tau,\vec{z},\vec{p},t_s)=C_{3pt}(\tau,t_s;\vec{z},\vec{p})/C_{2pt}(t_s;\vec{p})$ to take the advantage of the correlation between two-point and three-point functions. In the $t_s,\tau \rightarrow \infty$ limit, the ratio gives the ground-state matrix elements. In this work, we have calculated three time separation $t_s$ and done a two-state fit~\cite{Izubuchi:2019lyk} for the ground state extrapolation. In \fig{ratiofit}, we show ratios (data points) at $\vec{z}=\vec{0}$ of the two large momenta and the fitted results (colored bands). The black boxes are the ground state matrix elements and good agreement can be observed. In addition, the $|\vec{p}|$ = 2.24 GeV case shows smaller errors than the $|\vec{p}|$ = 2.15 GeV case especially for the ratio $R(\tau,\vec{z},\vec{p},t_s)$ at $t_s=8a$.

\begin{figure}
    \centering
    \includegraphics[width=0.4\textwidth]{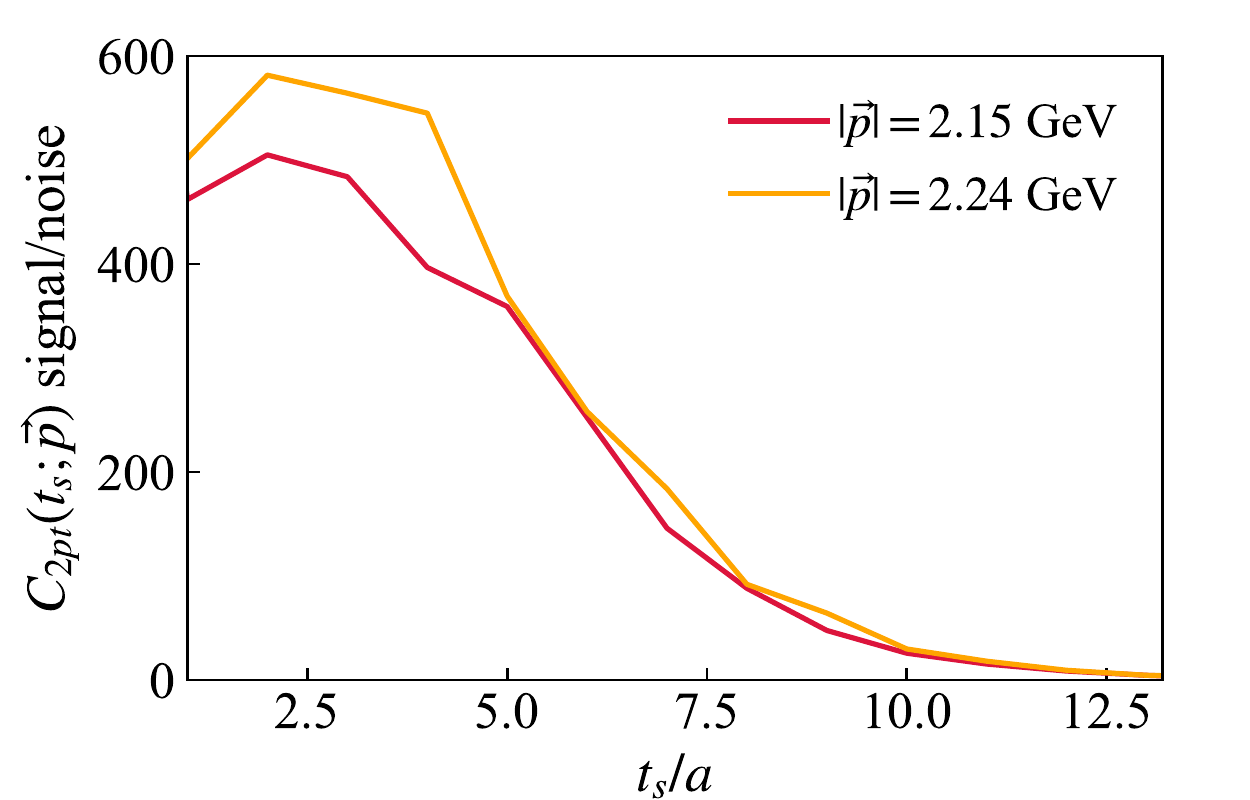}
    \includegraphics[width=0.4\textwidth]{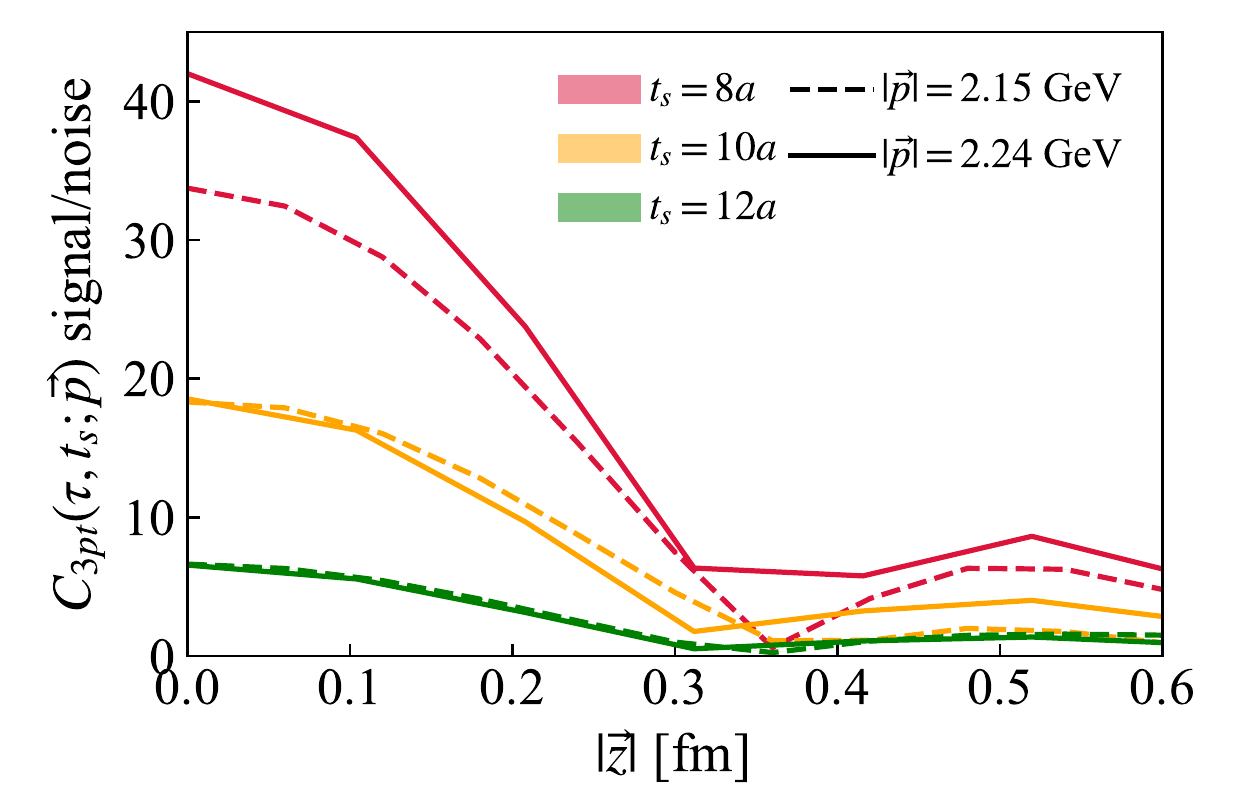}
    \caption{Upper panel: the signal-to-noise ratios of two-point functions $C_{2pt}(t_s;\vec{p})$ as a function of $t_s$ are shown. Lower panel: the signal-to-noise ratios of three-point functions $C_{3pt}(\tau,t_s;\vec{z},\vec{p})$ at $\tau=t_s/2$ are shown as a function of $z$ at different $t_s$.}
    \label{fig:SNR}
\end{figure}

To closely examine the signal difference between cases with $|\vec{p}|=2.15$ GeV ($\vec{n}=(0,0,5)$) and 2.24 GeV ($\vec{n}=(3,3,3)$), we present the signal-to-noise ratio (SNR) of their two-point functions as a function of $t_s$ in the upper panel of \fig{SNR}. At small values of $t_s$, the case with $|\vec{p}|$ = 2.24 GeV exhibits a higher SNR (approximately 1.2 times higher), suggesting that the off-axis data may have smaller statistical errors due to the presence of smaller momentum modes along each axis. As $t_s$ increases, both lines exhibit an exponential decrease proportional to $e^{-(E_p-m_\pi)t_s}$, as described 
in \textit{Lattice Methods for Quantum Chromodynamics} by DeGrand and DeTar~\cite{degrand2006lattice}.
Since the case with $|\vec{p}|$ = 2.24 GeV has slightly larger energy, its SNR decreases more rapidly, eventually becoming lower than the case with $|\vec{p}|$ = 2.15 GeV. A similar observation is made in the lower panel of \fig{SNR}, where we estimate the SNR of $C_{3pt}(\tau,t_s;\vec{z},\vec{p})$ with $\tau=t_s/2$ as a function of $z$. For smaller values of $t_s$, the off-axis case appears to perform better, while no advantage can be seen for large values of $t_s$ as they are dominated by the exponential decay of SNR from the larger energy.

\subsection{Discussion of the Gribov copies}\label{app:Gribov}

\begin{figure}
    \centering
    \includegraphics[width=0.4\textwidth]{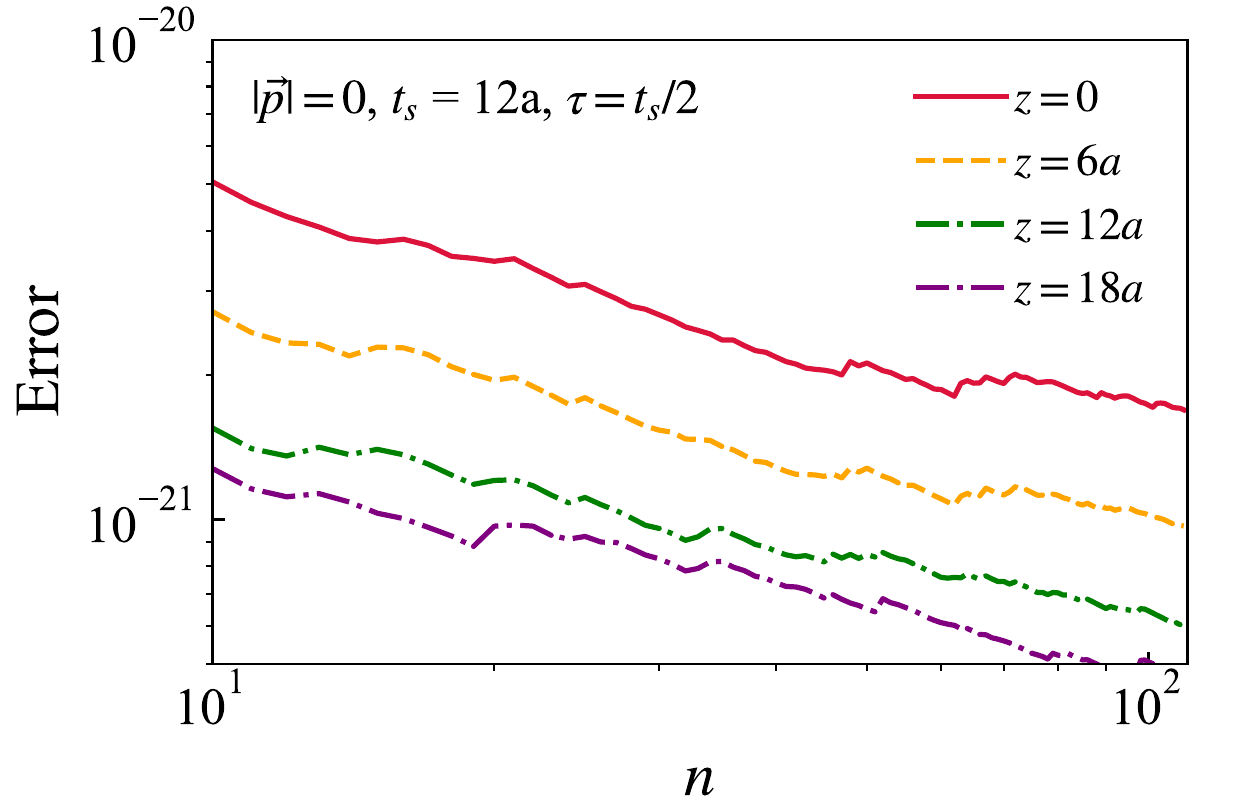}
    \includegraphics[width=0.4\textwidth]{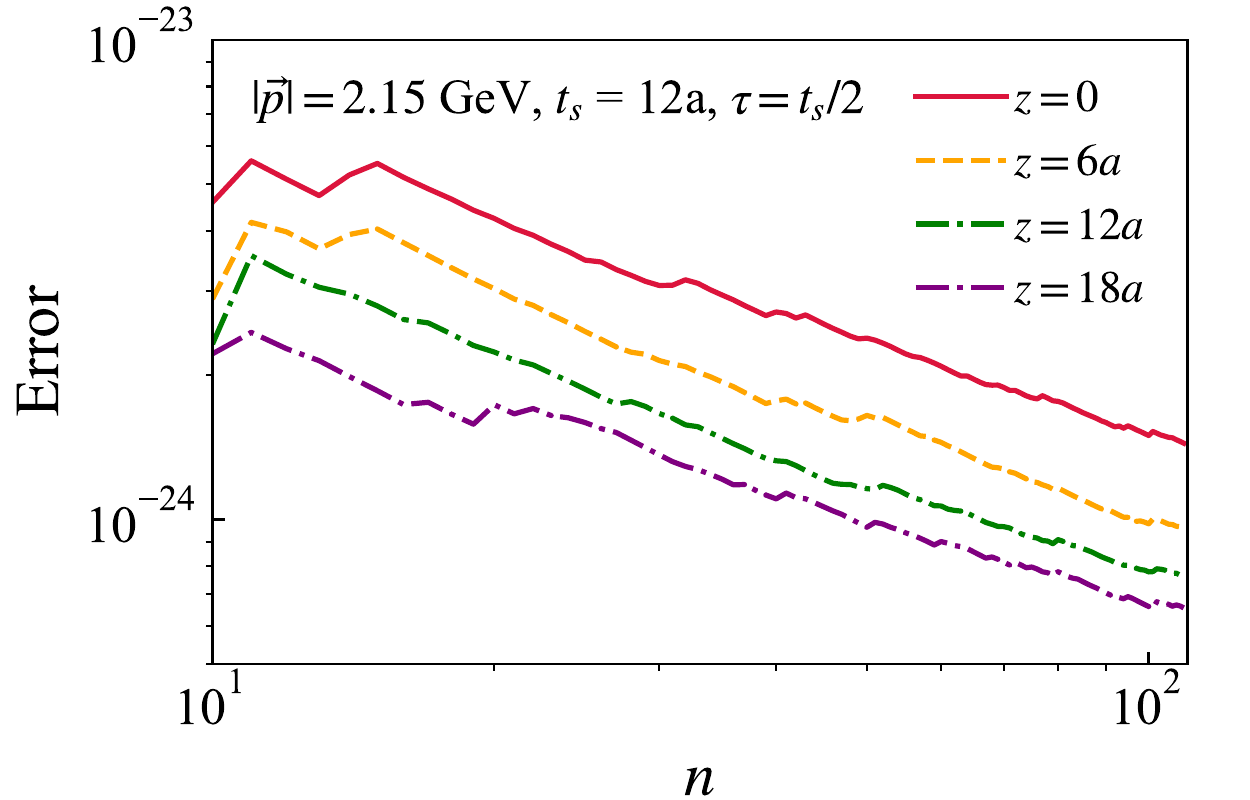}
    \caption{The errors of CG three-point functions $C_{3pt}(\tau,t_s;\vec{z},\vec{p})$ as a function of the number of configurations $n$ are shown for $|\vec{p}|=0$ (upper panel) and 2.15 GeV (lower panel).}
    \label{fig:Gribov}
\end{figure}

In non-perturabtive QCD, the gauge cannot be uniquely fixed due to the presence of Gribov copies~\cite{Gribov:1977wm,Singer:1978dk}. Consequently, gauge-dependent measurements obtained from different copies within the gauge orbit may exhibit some discrepancies. When performing configuration averaging, these systematic errors manifest as statistical errors that cannot be reduced even as the number of configurations increases. In \fig{Gribov}, we show the errors of CG three-point functions $C_{3pt}(\tau,t_s;\vec{z},\vec{p})$ at $t_s=12a$ and $\tau=t_s/2$, estimated from Jackknife methods, as a function of the number of configurations $n$. As one can see, up to all configurations we have, clear reduction in errors can be observed. What is more, the trend of $z\neq0$ matrix elements is the same as that at $z=0$, which is from a gauge-invariant local operator. In other words, the errors mainly originate from the statistical fluctuations, while the impact of the Gribov noise, if it exists, is negligible.

\subsection{Bare matrix elements and rotational symmetry}\label{app:bare2}

\begin{figure}
    \centering
    \includegraphics[width=0.4\textwidth]{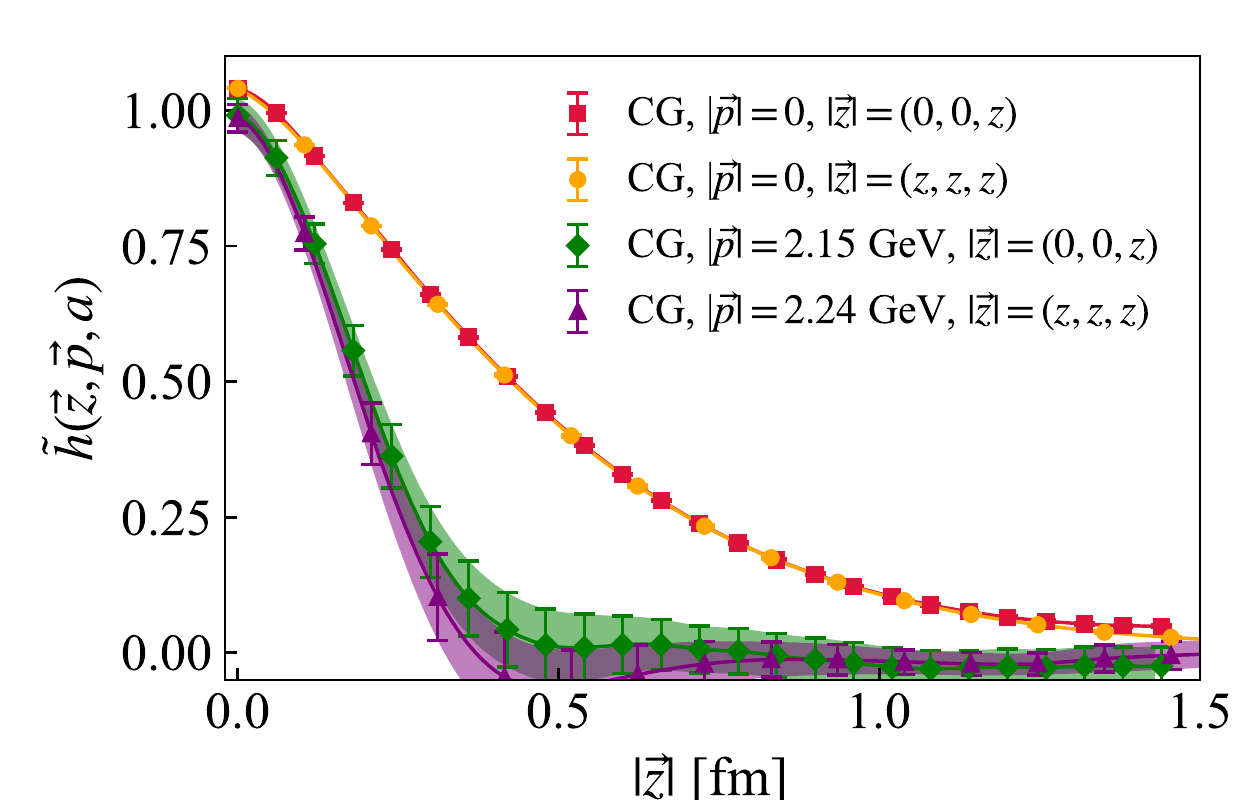}
    \caption{The bare matrix elements of CG qPDF as function of $|\vec{z}|$ from both on-axis and off-axis separations.}
    \label{fig:bmCG}
\end{figure}

\begin{figure}
    \centering
    \includegraphics[width=0.4\textwidth]{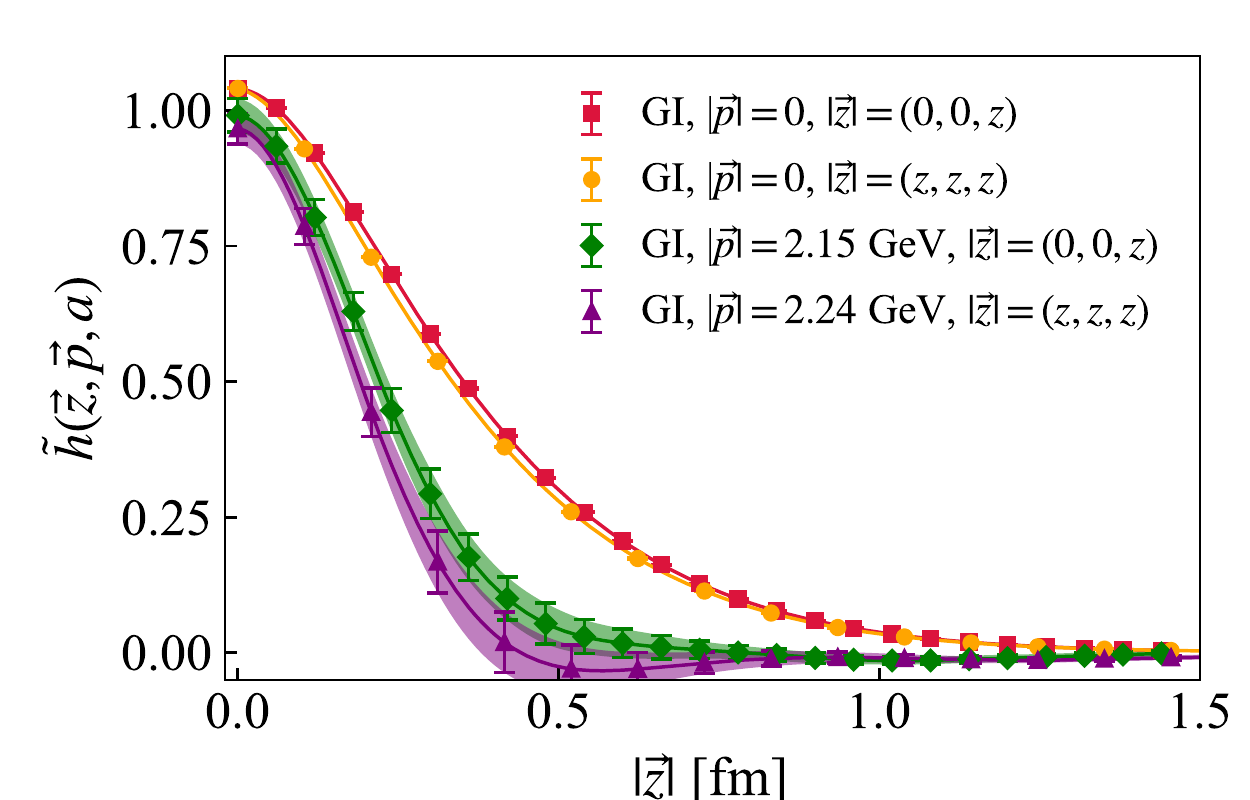}
    \includegraphics[width=0.4\textwidth]{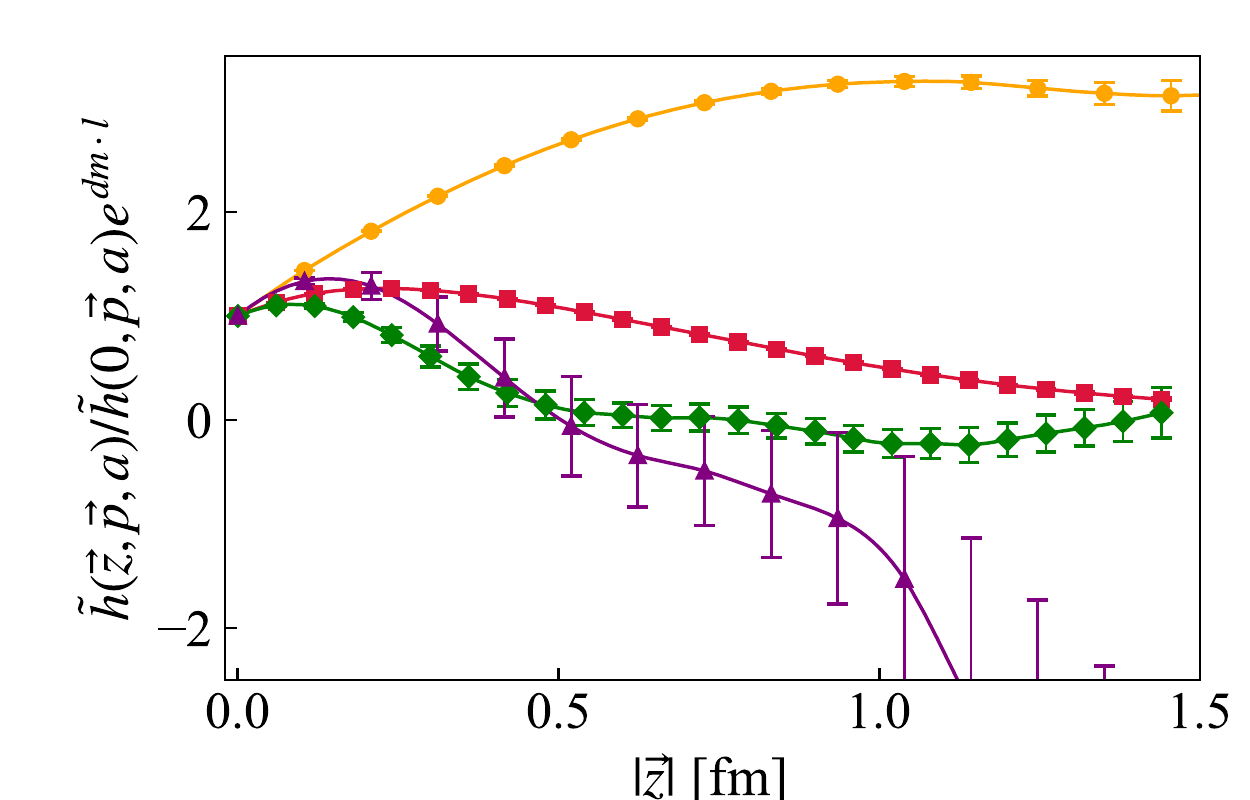}
    \caption{The bare (upper panel) and $\delta m$-subtracted matrix elements (lower panel) of GI qPDF.}
    \label{fig:bmGI}
\end{figure}

In \fig{bmCG}, we show the bare CG qPDF matrix elements for zero momentum as well as $|\vec{p}|=2.15$ GeV ($\vec{n}=(0,0,5)$) and $|\vec{p}|=2.24$ GeV ($\vec{n}=(3,3,3)$) as a function of $|\vec{z}|$. The direction of spatial separation, denoted as $\vec{z}$, aligns with the momentum vector. It is evident that the matrix elements from both on-axis and off-axis cases exhibit an excellent overlap, particularly at zero momentum, indicating the preservation of 3D rotational symmetry with high precision.

The zigzag Wilson line has been used to approximate the straight off-axis link as done in Ref.~\cite{Musch:2010ka} for the GI matrix elements. This approximation is based on the assumption that the measurements do not depend on the vector path but only the position. However this is certainly not true in lattice QCD as we all know the Wilson loops, e.g. plaquettes, are not trivially 1. The usage of HYP smearing can smear out some of the UV physics within the hypercube and make the plaquettes close to 1 (from $\sim$ 0.63 to $\sim$ 0.96 in our case). As a result, the matrix elements with zigzag Wilson line after HYP smearing are not significantly different from the straight Wilson line as shown in the upper panel of \fig{bmGI}. However, the several percent deviation is also obvious in contrast to the case of CG and It is hard to quantify this systematic error.

Also, we note that the total length of the zigzag Wilson line is $l=\sqrt{3}|\vec{z}|=3z$. Therefore, in the lower panel of \fig{bmGI} we show the matrix elements after subtracting the linear divergence $e^{-\delta m\cdot l}$, with $\delta m$ derived from the heavy quark potential ($\delta m  a=0.1586$)~\cite{Gao:2021dbh}. As one can see, $(e^{\delta m\cdot |\vec{z}|})^{\sqrt{3}}$ badly overshoots the linear divergence of matrix elements at off-axis $\vec{z}$, which makes their deviation from the on-axis $\vec{z}$ matrix elements even bigger. Again, the reason is that the HYP smearing distorted the UV physics within a hypercube and make the zigzag link close to a straight line. In summary, to use off-axis momenta with reasonable signal and rotational symmetry, the CG qPDF is the better choice.

\section{Renormalization}
\label{app:renorm}

\subsection{Ratio scheme}\label{app:sdf}

\begin{figure}
    \centering
    \includegraphics[width=0.9\columnwidth]{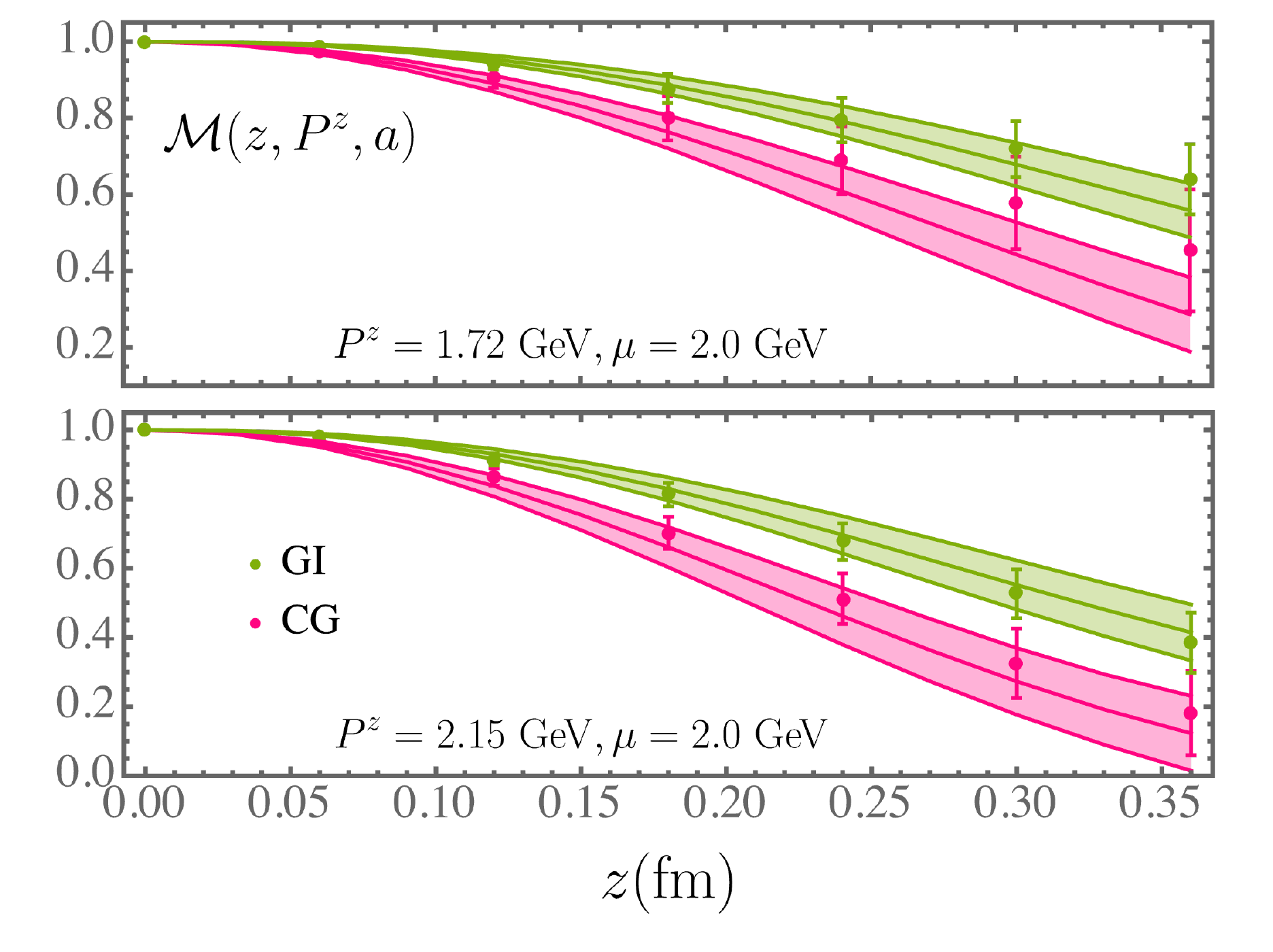}
    \caption{CG and GI ratios at $P^z=1.72$ and 2.15 GeV. The curves and bands are obtained by matching the PDF fitted from GI ratios at $n_z=4,5$ and $z\in[3a,6a]$ to the CG correlations and then taking their ratio.}
    \label{fig:sdf}
\end{figure}

With the bare matrix elements, we can check the consistency between CG and GI correlations at short distance. With a simple parametrizaton of the PDF, $f_v(x)\propto x^\alpha (1-x)^\beta$, we fit the ratio of GI correlations~\cite{Orginos:2017kos}
\begin{align}\label{eq:ratio}
	{\cal M}(z,P^z,a)&=\frac{\tilde h^{\rm GI}(z,P^z,a)}{\tilde h^{\rm GI}(z,0,a)}\frac{\tilde h^{\rm GI}(0,0,a)}{\tilde h^{\rm GI}(0,P^z,a)}\,,
\end{align}
at $z\in[3a, 6a]$ and $n_z=4,5$, with the NLO SDF formula. Then, we match the fitted PDF to the CG correlations using \eq{sdf}, calculate their ratio (pink band), and compare it to the lattice result (pink data points) in \fig{sdf}. After matching, the fitted PDF can describe the CG ratios within $1\sigma$ error, implying that the PDFs calculated from the CG and GI qPDFs must also be consistent at moderate $x$. The slight deviations could come from different ${\cal O}(z^2\Lambda_{\rm QCD}^2)$ corrections that are ignored in the SDF formulas or simply the statistical fluctuations.

\begin{figure}
    \centering
    \includegraphics[width=0.4\textwidth]{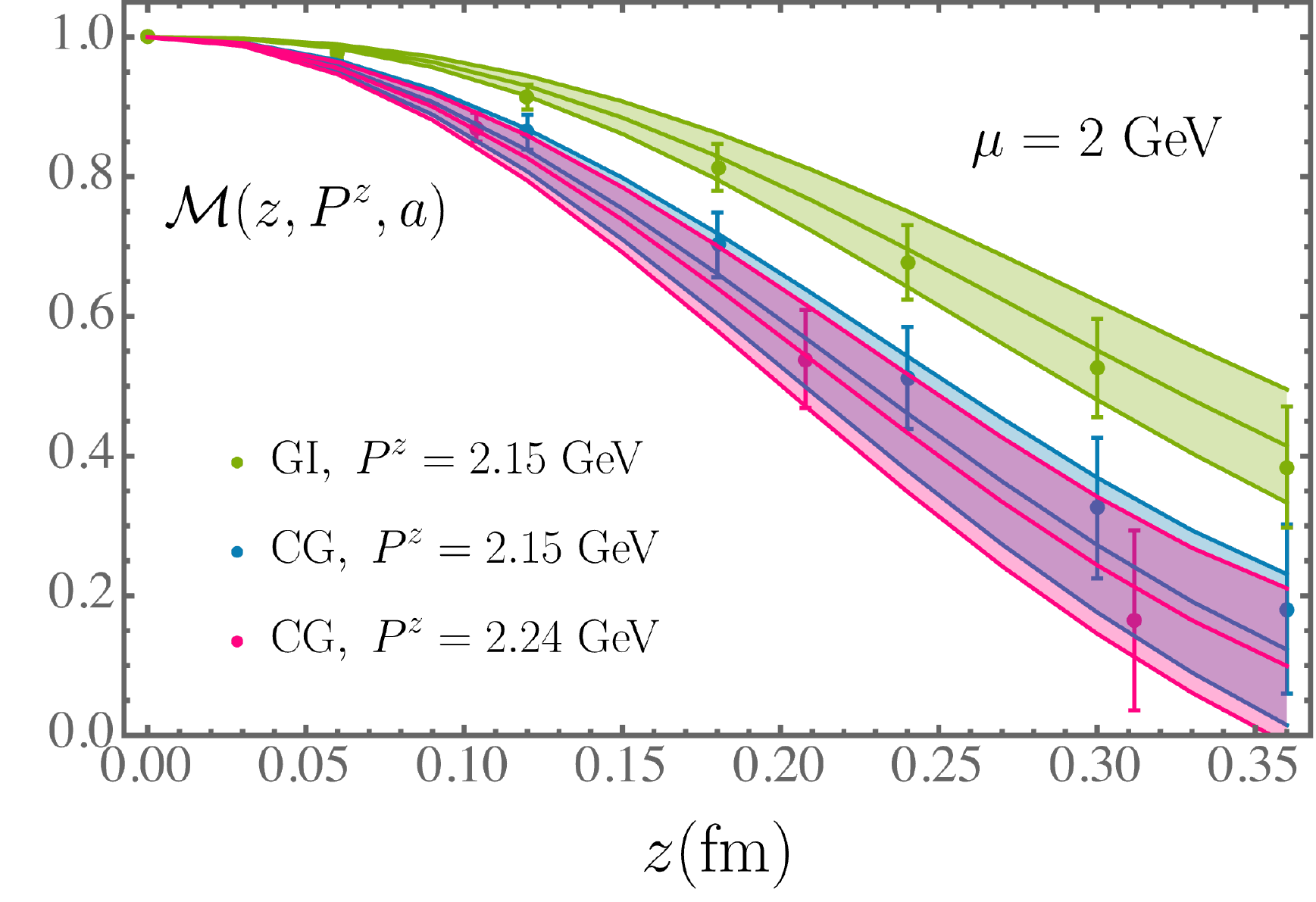}
    \caption{Comparison of CG and GI ratios with the inclusion of the off-axis momentum $|\vec{p}|=2.24$ GeV.}
    \label{fig:sdf2}
\end{figure}

In \fig{sdf2}, we compare the CG ratio at the off-axis momentum $|\vec{p}|=2.24$ GeV to the band predicted by matching the PDF fitted from the GI ratios. Like the $|\vec{p}|=2.15$ GeV case, the lattice ratio at $|\vec{p}|=2.24$ GeV agrees with the band within $1\sigma$ error, which is already implied by the rotational symmetry in \fig{bmCG}.

\subsection{Hybrid scheme}\label{app:hbd}

\begin{figure}
    \centering
    \includegraphics[width=0.4\textwidth]{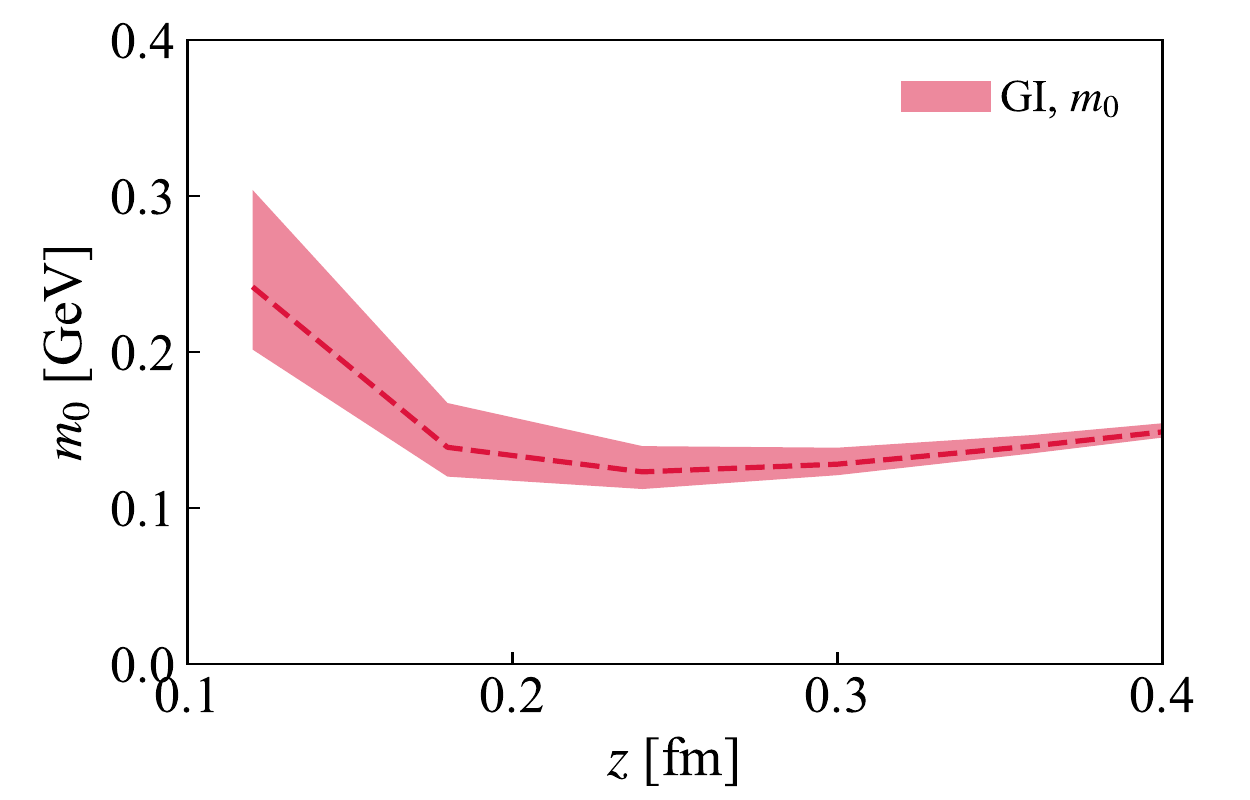}
    \caption{$m_0$ fitted with the leading-renormalon resummation (LRR) approach are shown as a function of $z$.}
    \label{fig:m0fit}
\end{figure}

\begin{figure}
    \centering
    \includegraphics[width=0.4\textwidth]{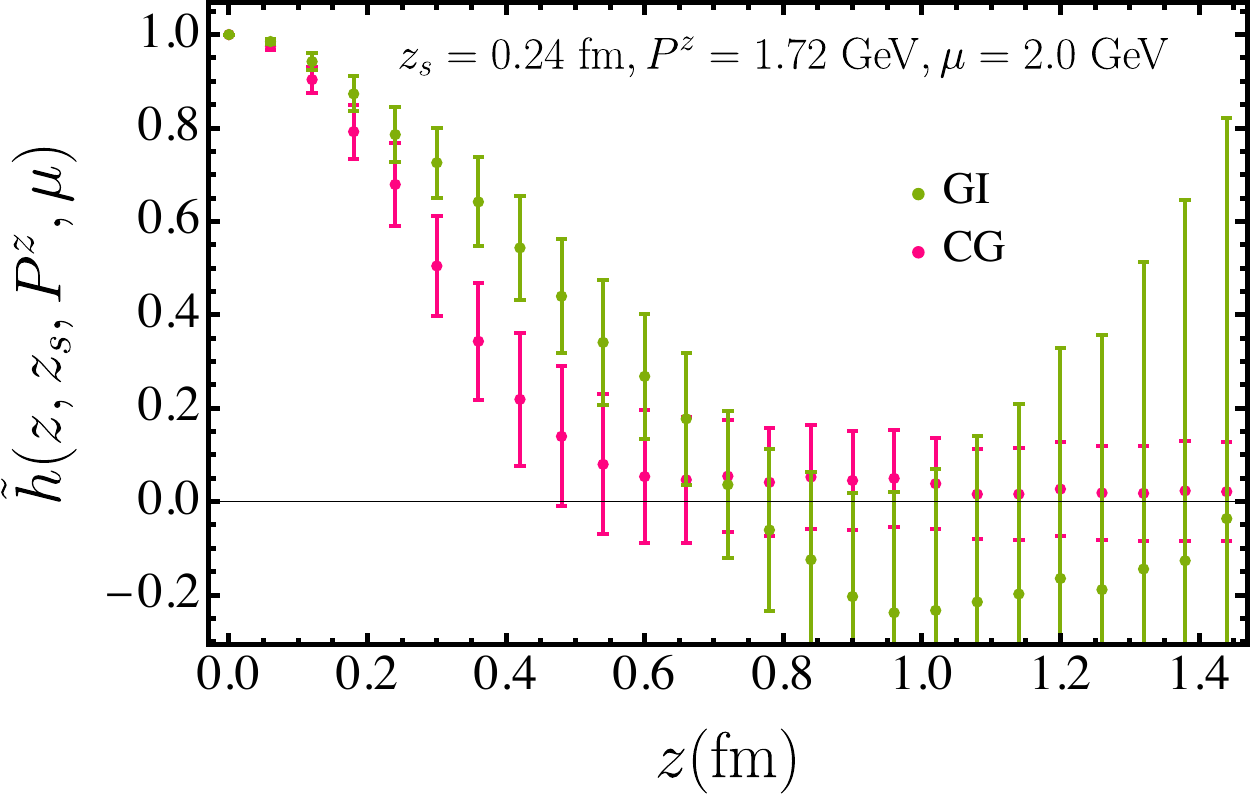}
    \caption{Hybrid-scheme CG and GI matrix elements at $P^z=1.72$ GeV.}
    \label{fig:hzpz4}
\end{figure}

We renormalize both the CG and GI matrix elements in the hybrid scheme defined in \eq{hbd}
As for the GI case, $\delta m$ is determined from the static quark-antiquark potential as $a\delta m=0.1586(8)$~\cite{Gao:2021dbh}. Then, we extract the linear renormalon $m_0(\mu)$ by comparing the $P_z=0$ matrix elements with the leading-renormalon resumed (LRR) Wilson coefficient under large-$\beta_0$ approximation~\cite{Holligan:2023rex,Zhang:2023bxs},
\begin{align}\label{eq:Effm0}
    e^{a \delta m}\frac{\tilde h(z,0,a)}{\tilde h(z-a,0,a)}=e^{-a m_0}\frac{C_0^{\rm LRR}(\mu^2z^2)}{C_0^{\rm LRR}(\mu^2(z-a)^2)} .
\end{align}
In \fig{m0fit}, we show $m_0(\mu)$ as a function of $z$ with the bands coming from the scale variation of $\mu\in[1.4,2.8]$ GeV. It can be seen that the $z$ dependence is mild except for the first point with serious discretization effect, suggesting the LRR Wilson coefficient plus the linear renormalon term can well describe the matrix elements at short distance. We take $m_0(2{\rm GeV})=0.1232(7)$ GeV at $z=0.24$ fm for the following analysis.

In \fig{hzpz4}, we compare the hybrid-scheme CG and GI qPDF matrix elements at $P^z=1.72$ GeV. Similar to \fig{hbd}, one can again observe more precise long-range correlations in the CG case.

\section{Matching}\label{app:match}

\subsection{NLO matching and NLL evolution}\label{app:match}

\begin{figure}
    \centering
    \includegraphics[width=0.4\textwidth]{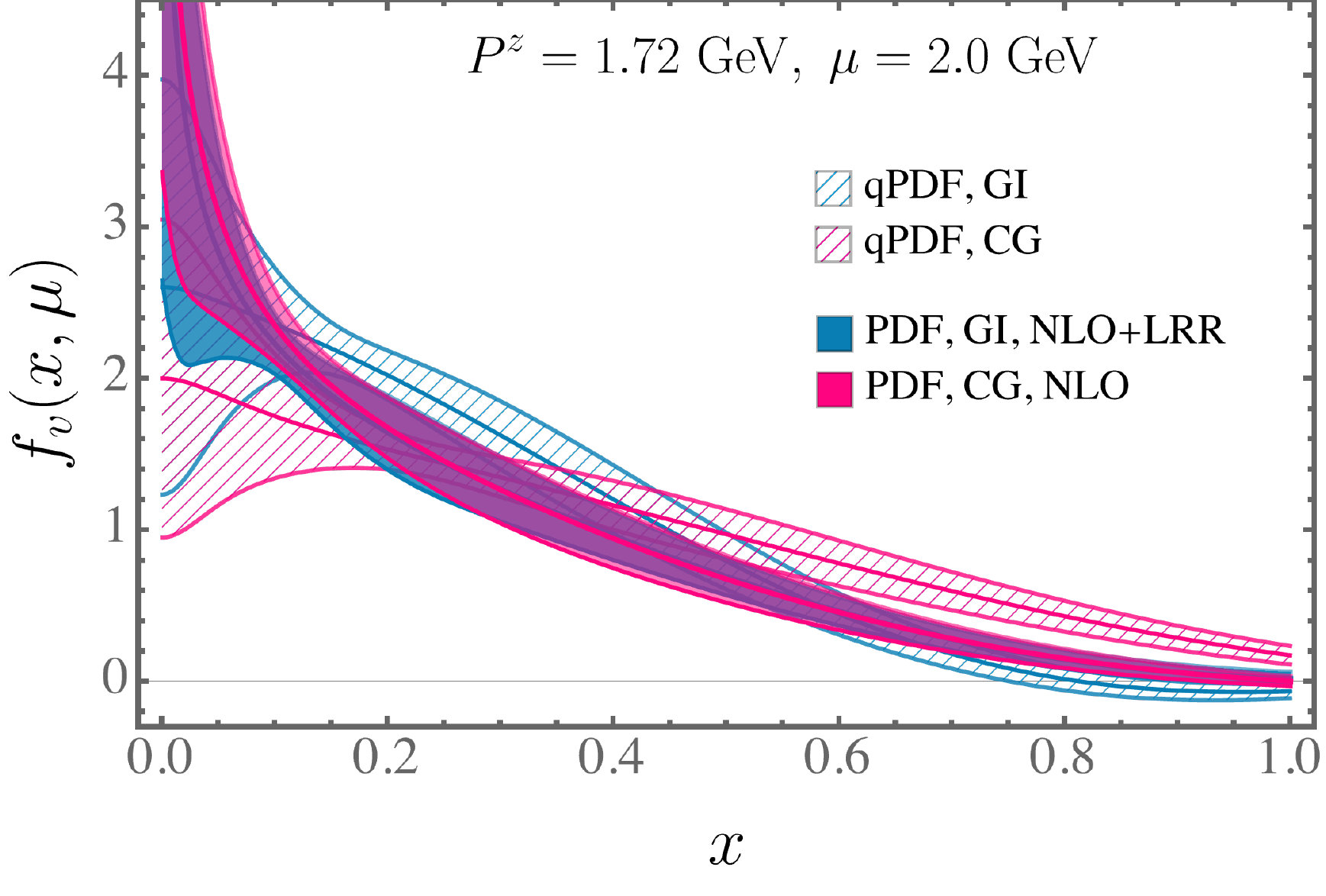}
    \caption{PDFs from the qPDFs after NLO matching at $P^z=1.72$ GeV.}
    \label{fig:match2}
\end{figure}

\begin{figure}
    \centering
    \includegraphics[width=0.4\textwidth]{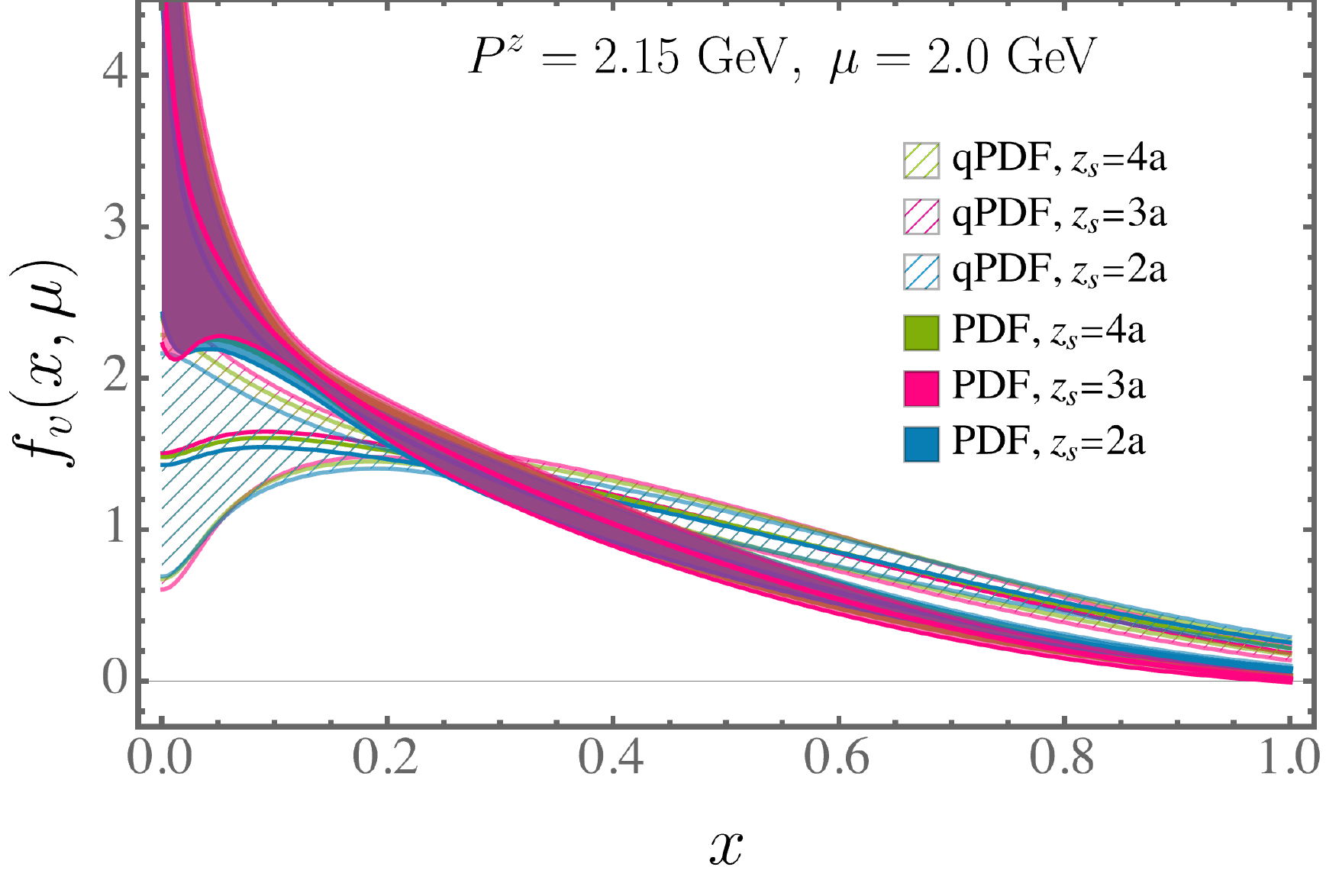}
    \caption{Comparison of the PDFs calculated from the CG qPDFs at $P^z=2.15$ GeV and different $z_s$.}
    \label{fig:zs}
\end{figure}

\begin{figure}
    \centering
    \includegraphics[width=0.4\textwidth]{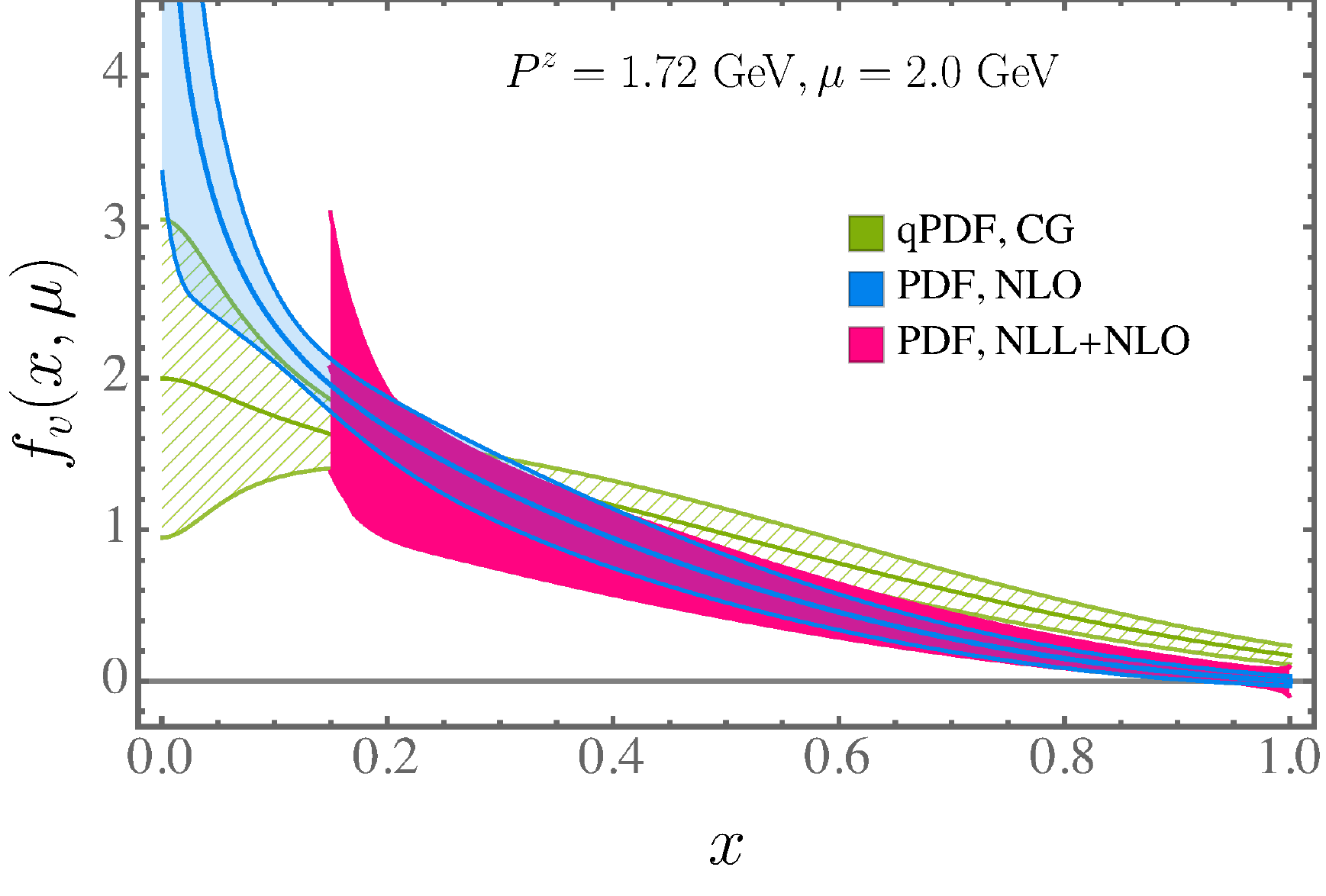}
    \includegraphics[width=0.4\textwidth]{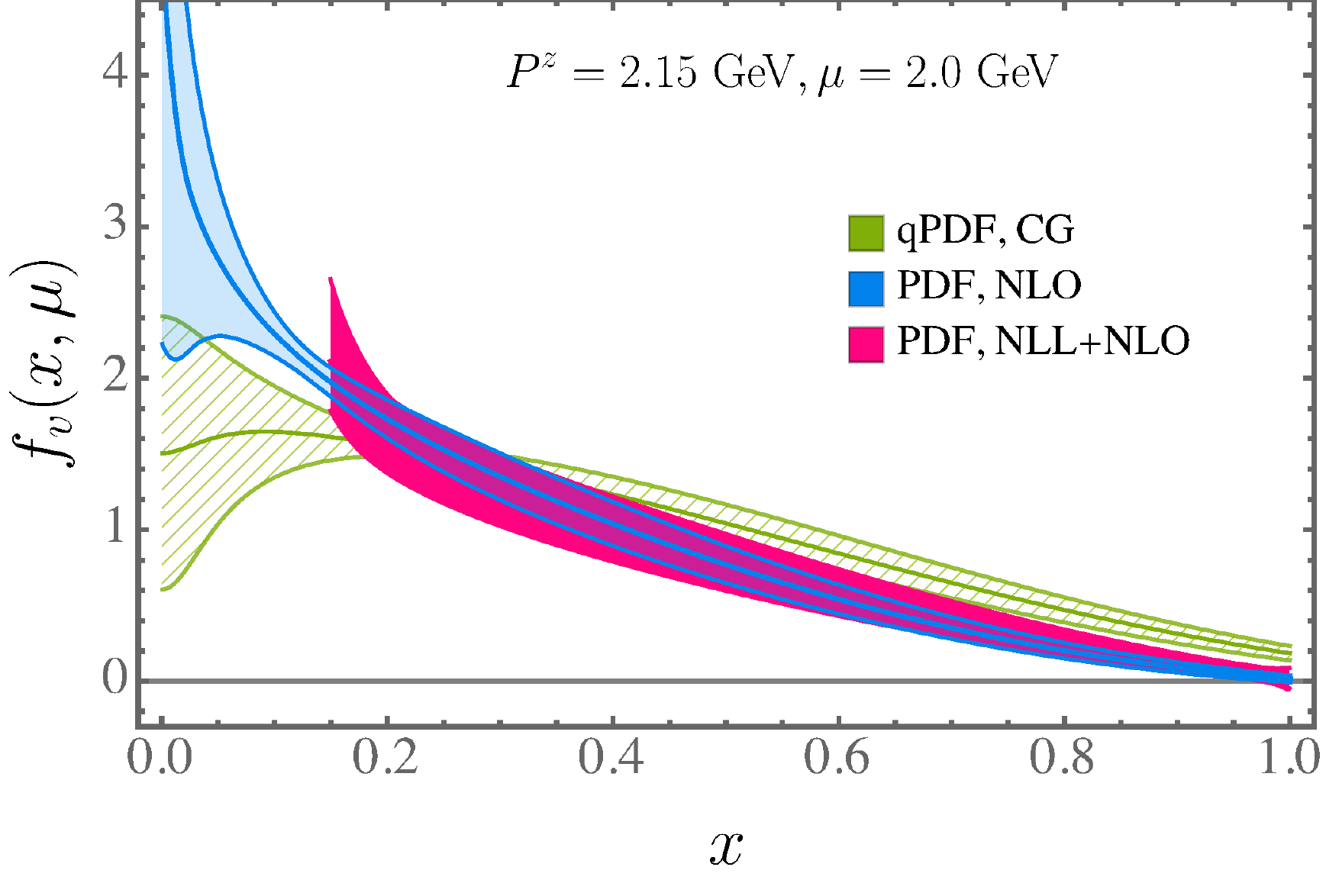}
    \includegraphics[width=0.4\textwidth]{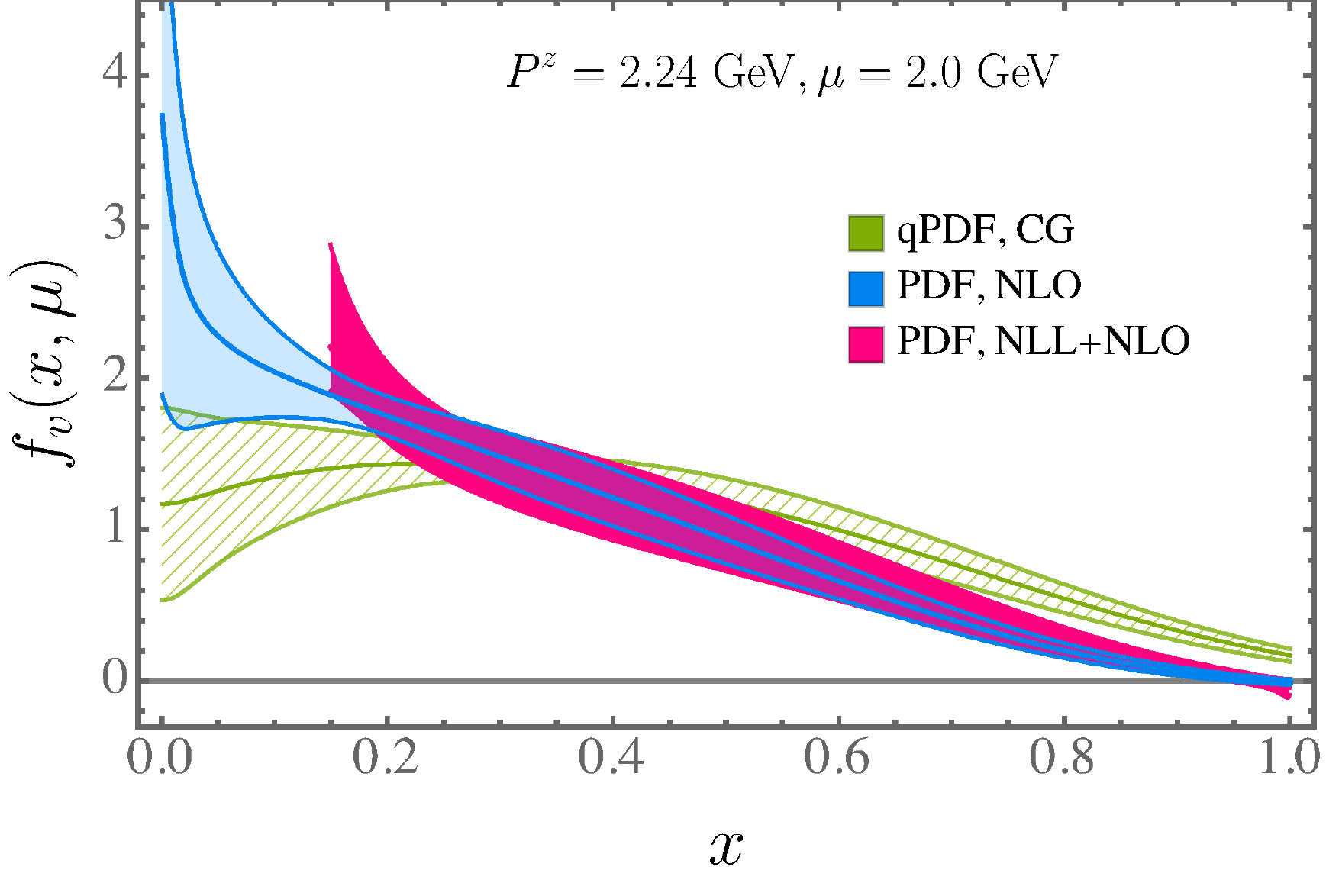}    
    \caption{PDFs matched from the CG qPDF at NLO and NLL+NLO.}
    \label{fig:NLOvsLL}
\end{figure}

\begin{figure}
    \centering
    \includegraphics[width=0.4\textwidth]{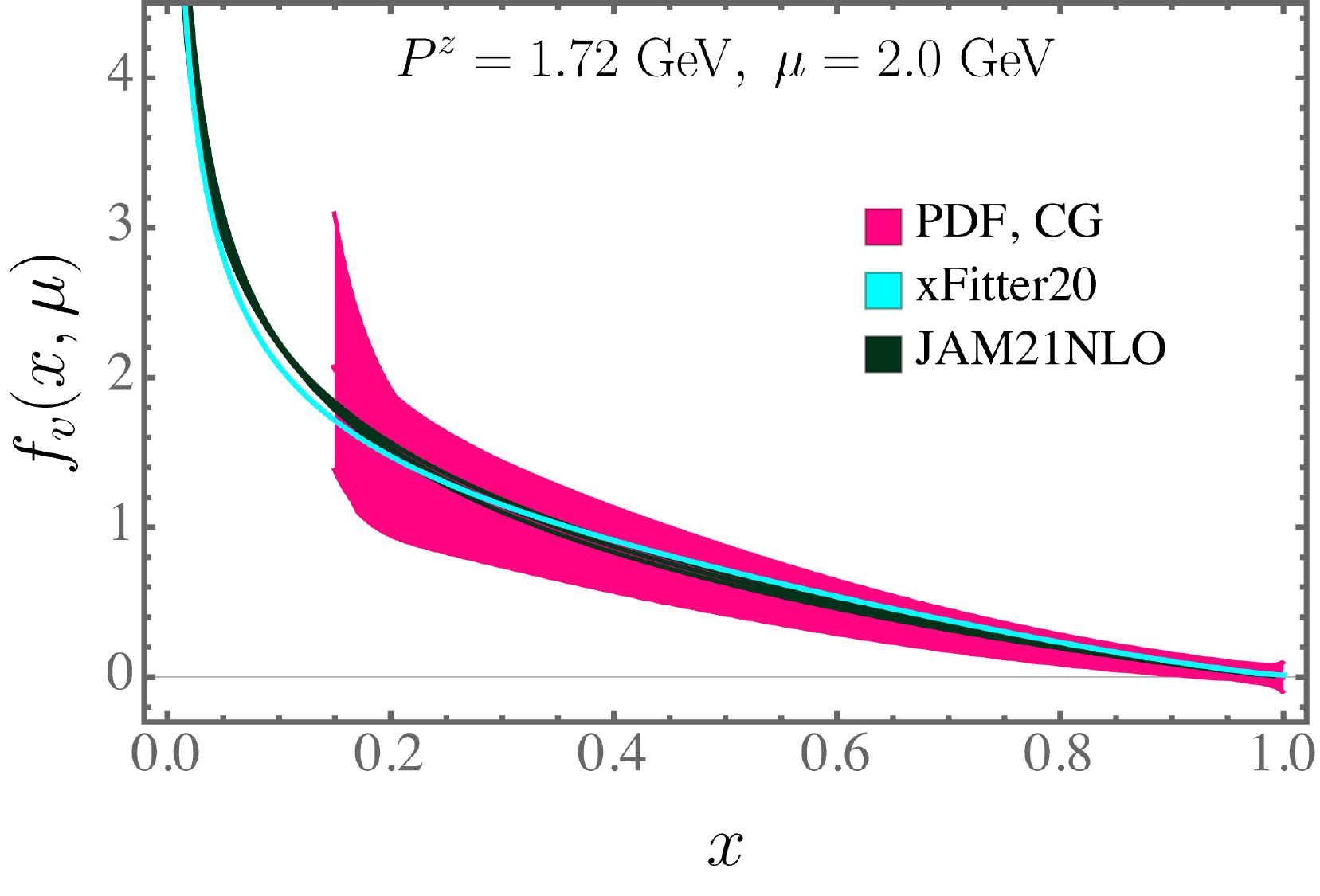}
    \caption{PDFs from CG qPDFs at $P^z=1.72$ GeV, compared to the global fits.}
    \label{fig:comp2}
\end{figure}

Following \fig{match}, we compare the PDFs calculated from the CG and GI qPDFs at  $P^z=1.72$ GeV at NLO in \fig{match2}. Again, despite the considerable differences between the CG and GI qPDFs, the matched PDFs show significantly improved agreement at moderate $x$.

Note that $z_s$ is an intermediate scale in the hybrid renormalization scheme. It must satisfy $a\ll z_s\ll \Lambda_{\rm QCD}^{-1}$ to avoid discretization and non-perturbative effects, and its dependence should be canceled by the matching correction. To verify this, we calculate the PDF from the CG qPDF at $P^z=2.15$ GeV and $z_s=\{2, 3, 4\}a$, which is shown in \fig{zs}. There is no noticeable difference among the results, except that $z_s=2a$ shows a little deviation from the others, which is mainly due to the discretization effects when $z_s\sim a$.

To demonstrate the effect of resumming small-$x$ logarithm or PDF evolution, we compare the PDFs matched from the CG qPDF at NLO and NLL+NLO accuracies in \fig{NLOvsLL}. For NLL resummation, we use two-loop evolution of $\alpha_s$~\cite{Gao:2021dbh}, which is defined by $\Lambda_{\overline{\rm MS}}=363.3$ MeV with the initial condition $\alpha_s(\mu=2.0{\rm\ GeV})=0.293$. The resummation has very little impact on the PDF at $x>0.4$, which mainly comes from the lower scale choice $\kappa=1/\sqrt{2}$, but becomes more and more significant as $x$ decreases. Eventually, at $2xP^z \sim 0.8$ GeV where $\alpha_s$ becomes of ${\cal O}(1)$, the scale variation uncertainty becomes out of control.

Finally, for completeness we include a comparison of the PDF calculated from the CG qPDF at $P^z=1.72$ GeV to the global fits in \fig{comp2}. Again we find agreement between lattice and phenomenology at moderate to large $x$, whose central value aligns slightly closer than the two larger momenta cases. Since the statistical errors in the current lattice results are not small, the unquantified power corrections, which should be better suppressed at higher momenta, may just be a less important systematic uncertainty here.

\subsection{Convergence of the GI and CG methods}
\label{app:conv}

\begin{figure}
    \centering
    \includegraphics[width=0.4\textwidth]{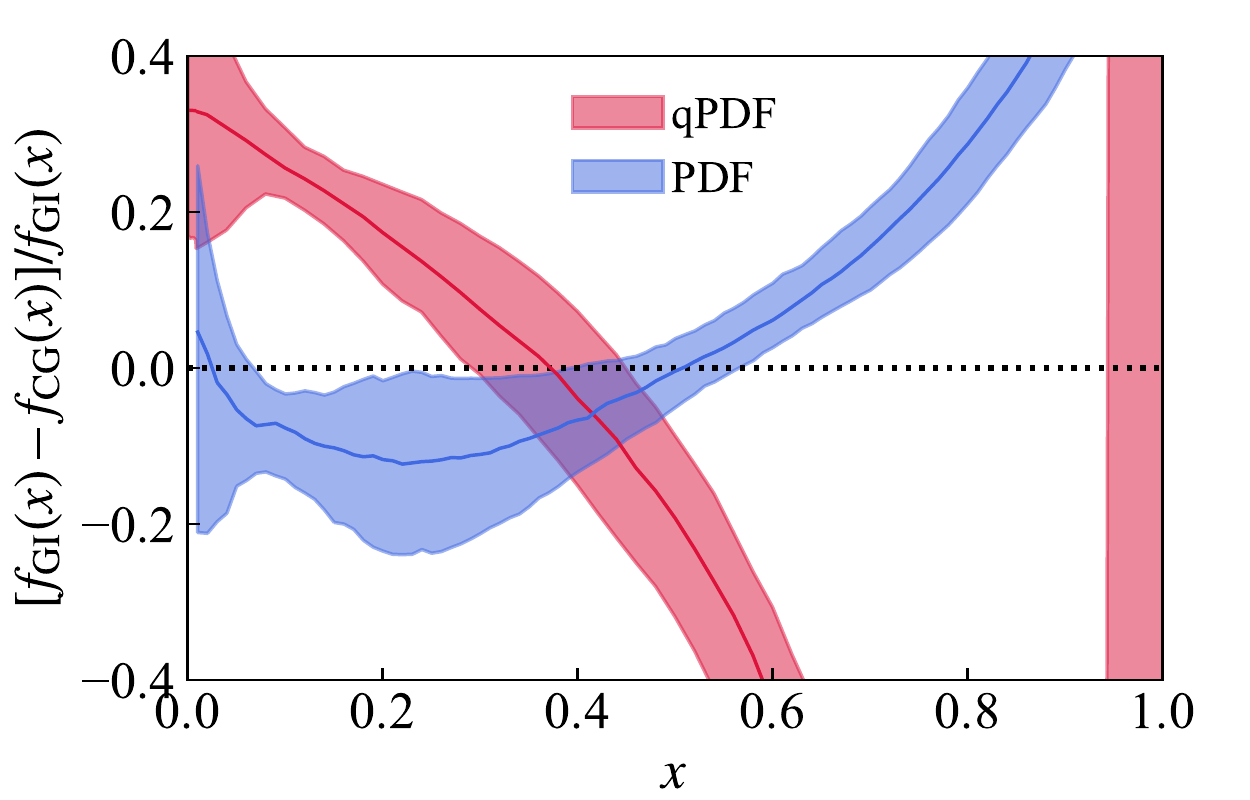}
    \caption{The relative difference $D_f(x)$ between the GI and CG quasi-PDFs (red band) and matched PDFs (blue band).}
    \label{fig:diff}
\end{figure}

\begin{figure}
    \centering
    \includegraphics[width=0.5\textwidth]{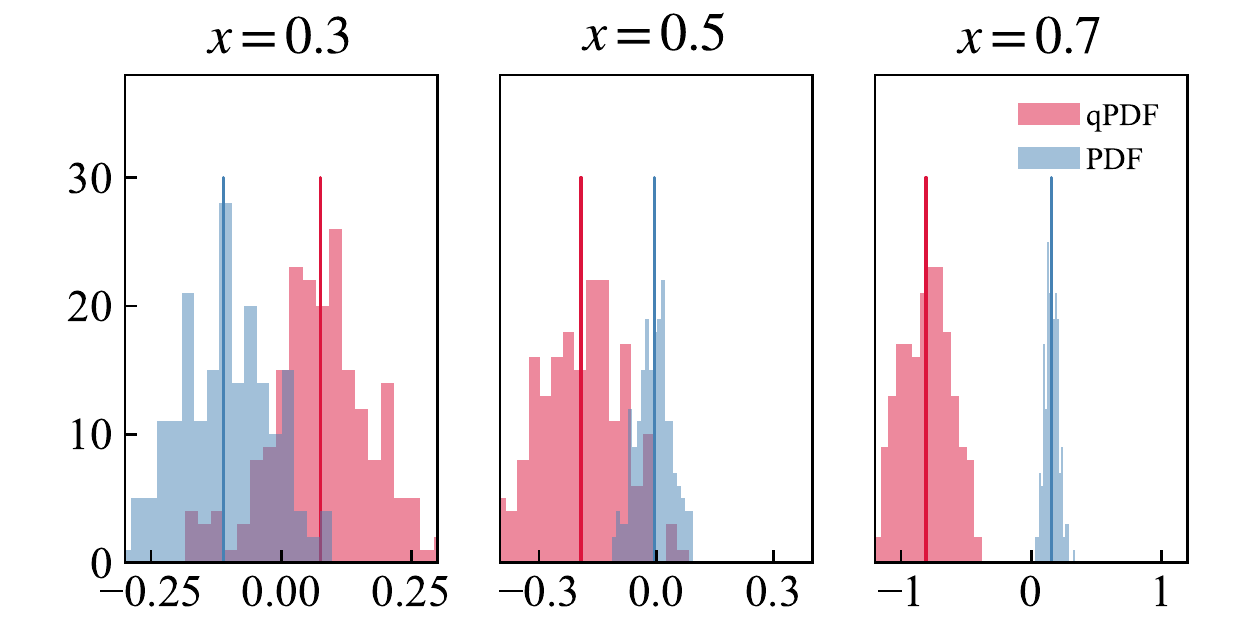}
    \caption{The distribution of the bootstrap samples of $D_f(x)$ for quasi-PDF (red histograms) and PDF (blue histograms) at $x=0.3,0.5,0.7$.}
    \label{fig:diffx}
\end{figure}

As mentioned in the main text, a distinction has been observed between the CG and GI qPDFs. However, following perturbative matching, the resulting PDFs demonstrate a remarkable level of agreement, particularly in the moderate $x$ region. For a more direct observation, while also considering the correlation between GI and CG matrix elements, we calculate the relative difference between them sample by sample,
\begin{align}
    D_f(x)=\frac{f_{\rm GI}(x)-f_{\rm CG}(x)}{f_{\rm GI}(x)}\,.
\end{align}
The $D_f(x)$ derived from qPDF (red band) and PDF (blue band) are shown in \fig{diff} using the same data as \fig{match}. Consistent with the observation in \fig{match}, we find excellent agreement in the PDFs across a wide range of moderate $x$, despite substantial difference in the quasi-PDFs. In \fig{diffx}, we present the distributions of bootstrap samples of $D_f(x)$ at $x=0.3, 0.5, 0.7$ for the qPDFs (red histograms) and PDFs (blue histograms). Notably, both the distribution and central value of $D_f(x)$ from the PDFs tend to converge toward zero. This further demonstrates that the CG and GI qPDFs belong to the same universality class for calculating the PDF.

\newpage

\bibliography{xpdf}

\end{document}